\documentclass[sigconf]{acmart}

\AtBeginDocument{%
}

\copyrightyear{2025}
\acmYear{2025}
\acmConference[UIST '25]{The 38th Annual ACM Symposium on User Interface Software and Technology}{September 28-October 1, 2025}{Busan, Republic of Korea}
\acmBooktitle{The 38th Annual ACM Symposium on User Interface Software and Technology (UIST '25), September 28-October 1, 2025, Busan, Republic of Korea}\acmDOI{10.1145/3746059.3747632}
\acmISBN{979-8-4007-2037-6/2025/09}

\usepackage{acmart-taps}  
\usepackage{amsmath}  
\newcommand{\Z}{\mathbb{Z}}
\newcommand{\D}{\ensuremath{\Delta}}

\usepackage{algorithm}
\usepackage{algpseudocode}

\newcommand{\figref}[1]{Fig.~\ref{#1}}

\newcommand{\secref}[1]{Sec.~\ref{#1}}
\newcommand{\appref}[1]{App.~\ref{#1}}

\usepackage{graphicx}   
\usepackage{wrapfig}    
\usepackage{float}
\usepackage{dblfloatfix}

\usepackage{enumerate}
\usepackage{enumitem}

\newlist{goals}{enumerate}{1}
\setlist[goals]{
  label=(G\arabic*),    
  ref=G\arabic*         
}

\newlist{constraints}{enumerate}{1}
\setlist[constraints]{
  label=(C\arabic*),
  ref=C\arabic*
}

\usepackage{booktabs}   
\usepackage{siunitx}    

\setlist{leftmargin=*}  
\setlist{itemsep=0pt}   
\setlist{topsep=0pt}    
\setlist{parsep=0pt}    

\usepackage{xspace}
\newcommand{\garmod}{\textsc{Refashion}\xspace}

\newcommand{\defn}[1]{\textbf{\textit{#1}}}

\newcommand{\edited}[1]{#1}

\newcommand{\participantquote}[1]{\textit{``#1''}}

\begin{document}

\title{\garmod: Reconfigurable Garments via Modular Design}

\author{Rebecca Lin}
\email{ryelin@mit.edu}
\affiliation{
  \institution{MIT CSAIL}
  \city{Cambridge}
  \state{MA}
  \country{USA}
}

\author{Michal Lukáč}
\email{lukac@adobe.com}
\affiliation{
  \institution{Adobe Research}
  \city{San Jose}
  \state{CA}
  \country{USA}
}

\author{Mackenzie Leake}
\email{leake@adobe.com}
\affiliation{
  \institution{Adobe Research}
  \city{San Francisco}
  \state{CA}
  \country{USA}
}

\begin{abstract}
While bodies change over time and trends vary, most store-bought clothing comes in fixed sizes and styles and fails to adapt to these changes. Alterations can enable small changes to otherwise static garments, but these changes often require sewing and are non-reversible. We propose a modular approach to garment design that considers resizing, restyling, and reuse earlier in the design process. Our contributions include a compact set of modules and connectors that form the building blocks of modular garments, a method to decompose a garment into modules via integer linear programming, and a digital design tool that supports modular garment design and simulation. Our user evaluation suggests that our approach to modular design can support the creation of a wide range of garments and can help users transform them across sizes and styles while reusing the same building blocks.
\end{abstract}



\begin{teaserfigure}
\centering
\includegraphics[width=\textwidth]{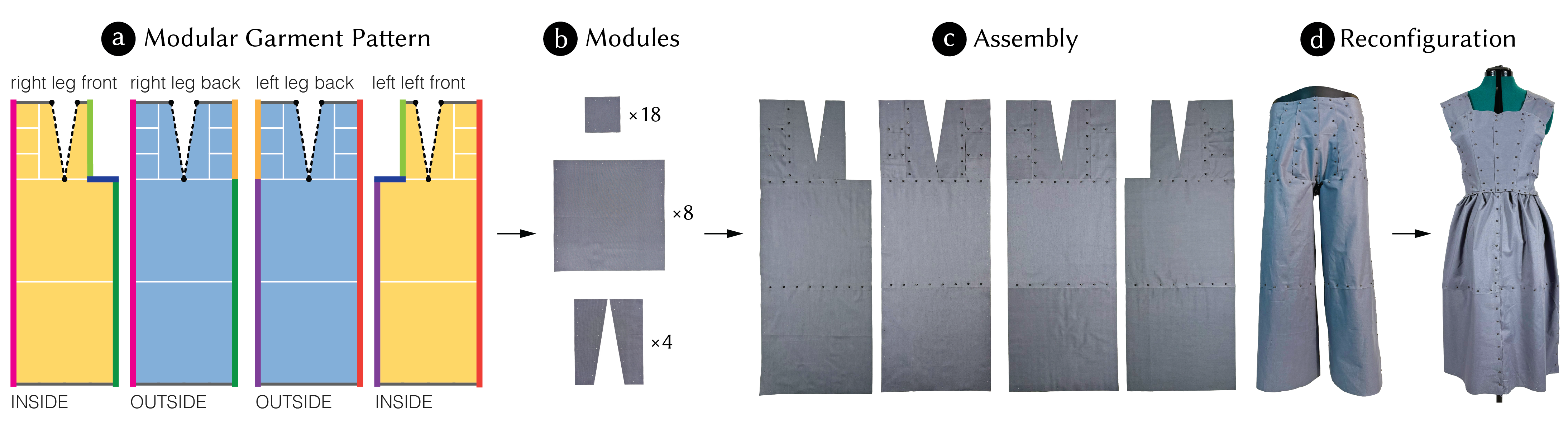}
\caption{\label{fig:teaser}
We introduce a system for designing modular garment patterns (a) that can be constructed from a shared set of fabric modules (b) into complete garments (c). These garments—such as the trousers shown—can be reconfigured to adjust fit and style, or even fully transformed into a different garment, such as the dress (d), all while reusing the same set of building blocks.}
\Description{Modular garments demonstrating full reconfiguration. Depicts a modular garment pattern and its reassembly. (A) Pattern of square and trapezoid modules. (B) Two sizes of square modules and one pair of trapezoids. (C) Modules laid flat and then draped as trousers on a mannequin. (D) The same modules reassembled into a sleeveless flared dress.}
\end{teaserfigure}

\maketitle

\section{Introduction}\label{sec:introduction}

Most commercial garments are produced in a limited range of sizes and styles to streamline manufacturing. Individuals with sewing skills can adjust certain aspects of fit, such as shortening pant legs or reducing a skirt’s volume, or perform simple repairs, such as patching a hole in jeans. However, these alterations remain difficult for the average consumer and often result in irreversible changes. For example, once pants are cut into shorts, restoring them without visible signs of alteration is nearly impossible.

While tailoring and mending play important roles in improving garment fit and lifespan, there are advantages to considering adaptability and reusability earlier in the garment design process.
Rather than encouraging consumers to discard garments that no longer fit or suit their style, designing for easy resizing, restyling, and repair promotes longevity.
Inspired by recent work in zero-waste garment design~\cite{zhang2024wastebanned} and design for disassembly and reconfigurability in textiles and other domains~\cite{song2017reconfigurable,wu2020unfabricate,wen2025enabling}, we propose a modular approach to fashion. This approach encourages the design of garment elements intended to be reconfigured across fits and styles.
Modular design involves constructing systems from independent and often interchangeable \emph{modules} connected through well-defined \emph{interfaces}.
This design paradigm has enabled scalability and reuse in many domains such as furniture design~\cite{koo2016towards, yao2017interactive, wu2019carpentry}, architecture and construction~\cite{eigensatz2010paneling}, consumer electronics, and software.
In this paper, we consider how these principles can be applied to fashion to facilitate the creation of dynamic garment systems.

Reconfigurable and resizable garments have emerged as a growing area of interest in fashion through small-scale explorations and limited-run clothing lines. For example, Sofia Ilmonen~\cite{sofiailmonen} introduced expertly crafted apparel comprising uniform fabric modules connected with buttons and loops.
Petit Pli~\cite{petitpli} developed origami-inspired children's wear that grows with the child.
Other commercially available reconfigurable garments feature detachable components~\cite{marfastance}, such as sleeves or hoods, or employ mix-and-match approaches, such as treating tops and bottoms as interchangeable modules~\cite{larocca}. However, such systems offer limited versatility and user customizability beyond simple variations of a base garment or a small set of predefined pairings.

While there are many possible interpretations of modularity in garment design, we identify three principles to characterize this design space: First, the modules should be \emph{reusable} across a diverse range of garment designs and users. Second, the interfaces must support \emph{reversible} yet secure connections to enable repeated assembly while ensuring wearer comfort. Lastly, both the modules and the resulting garments should be easy for users to fabricate and assemble. In response to these goals, we introduce \garmod, a new system for modular garment design. Our contributions include:
\begin{enumerate}
\item \textbf{Modules:} A compact, reusable set of modules that act as the building blocks for modular garments. Although they are standardized to reduce fabrication waste and promote easy assembly, these modules support expressive shaping techniques such as gathers, pleats, and darts (\secref{sec:modules}).
\item \textbf{Decomposition Algorithm:} An optimization-based approach that uses integer linear programming to decompose garment patterns into a minimal set of modules for efficient assembly (\secref{sec:assembly_formulation}).
\item \textbf{Modular Design Tool:} A digital tool for creating, manipulating, and simulating modular garment patterns—enabling the exploration of design possibilities alongside their corresponding assembly instructions (\secref{sec:system}).
\end{enumerate}
Our results and user studies indicate that this modular approach to garment design supports the creation of many different garment types that can be resized, restyled, and repurposed from a shared set of core building blocks.
\section{Related Work}\label{sec:related}

We build upon prior work in computational garment design, sustainable fabrication, and designing assemblies that support modularity.

\subsection{Computational Garment Design Tools}
Most garments are constructed by cutting, folding, and stitching 2D panels into 3D forms. Several computational tools have been developed to bridge the gap between 2D sewing patterns and 3D garment models—through simulation~\shortcite{clo3D}, bi-directional editing between formats~\shortcite{umetani2011sensitive, berthouzoz2013parsing, bartle2016physics, qi2024perfecttailor}, and automatic pattern inference from 3D shapes~\cite{liu20183d, pietroni2022computational}.
For garment shaping, prior work has investigated the placement of pleats~\cite{li2018foldsketch} and darts~\cite{de2023perfectdart}, as well as the alignment of fabric prints across seams~\cite{wolff2019wallpaper}.
Other tools have focused on garment sizing.
While commercial workflows typically rely on standardized ``size grading'' techniques, recent approaches explore more personalized garment retargeting using optimization~\cite{wolff2023designing} and parameterized patterns for custom body measurements~\shortcite{korosteleva2023garmentcode, freesewing}.
Closest to modular garments is GarmentCode~\shortcite{korosteleva2023garmentcode}, which treats semantic garment components (e.g., sleeves, skirts, etc.) as swappable modules. However, such systems are typically constrained to specific garment types, making it difficult to repurpose modules across distinct categories (e.g., reusing parts of a skirt in a blouse). In contrast, we introduce a smaller set of more universal building blocks that can be combined to produce multiple types of garments.

\subsection{Reducing Waste in Fabrication}

Prior work in computational design and fabrication has explored a range of strategies to promote material-efficient practices. One common approach is enabling reuse of existing materials. For example, Scrappy~\shortcite{wall2021scrappy} supports the reuse of leftover 3D printing materials as infill, while EcoEDA~\shortcite{lu2023ecoeda} facilitates the design of electronic circuits using recycled components by offering reuse suggestions and maintaining a component library. Other efforts have focused on biodegradable alternatives, such as heating pouches made from leaves~\shortcite{song2022towards}, filament derived from used coffee grounds~\shortcite{rivera2023designing}, and compostable threads for textiles~\shortcite{zhu2024ecothreads}. Another area of focus has been integrating material awareness early in the design process. Fabricaide~\shortcite{sethapakdi2021fabricaide}, for instance, helps users optimize material sheet usage during laser cutting. Similarly, Koo et al.~\shortcite{koo2016towards} and Wu et al.~\shortcite{wu2019carpentry} introduced tools for designing low-waste furniture layouts from flat sheets. Beyond specific tools, broader HCI research has examined themes such as reuse, salvage fabrication, and the practice of unmaking as part of sustainable design discourse~\shortcite{song2021unmaking, dew2018making, jones2021computational}.

Within the context of textiles, several systems have aimed to reduce waste and encourage reuse. Visible mending techniques have been explored as an aesthetically intentional form of repair~\shortcite{jones2021patching}, and ScrapMap~\cite{leake2024scrapmap} helps users incorporate fabric scraps into quilting projects. Knit and crochet textiles offer an inherently low-waste fabrication method, using only the necessary yarn. Recent work texTile~\cite{del2025textile} extends this efficiency by considering a modular approach to granny square garments—akin to our own—using custom connectors to support reuse and reconfiguration. In contrast, cut-and-sew garments often produce more waste due to the irregular shapes of fabric panels, which leave behind awkward scraps. InStitches~\shortcite{leake2023institches} helps users repurpose these remnants to practice sewing techniques. The field of zero-waste fashion design~\cite{rissanen2016zero} takes this challenge further by encouraging the creation of clothing that eliminates fabric waste entirely. WasteBanned~\cite{zhang2024wastebanned} supports users in adhering to these constraints by providing tools that consider material usage throughout the design process. Our system contributes to this space by representing garments as assemblies of rectangular modules, which enable highly efficient packing on fabric sheets and minimize textile waste.

\subsection{Designing Reusable Assemblies}
Efficient physical assembly has been a long-standing goal in computational design, with strategies aimed at promoting material reuse, streamlining manufacturing, and reducing construction time.
One foundational approach is shape decomposition—dividing a shape into meaningful subcomponents~\cite{shamir2008survey}. A well-known variant aimed at maximizing reuse is geometric dissection, which seeks to transform one shape into another of equal area. Because this problem is NP-hard~\cite{bosboom2015k}, researchers have proposed approximate methods that tolerate small discrepancies~\cite{duncan2017approximate}. These solutions often account for practical assembly constraints, such as reversibility, by requiring connections that are secure yet non-permanent~\cite{song2017reconfigurable, li2018construction}.
We extend these ideas to garments by decomposing clothing into modular components that can be disassembled and reconfigured into new designs.

\edited{Another important design goal is minimizing part diversity to streamline fabrication. In architecture, for example, freeform surfaces are often constructed from a small number of standardized panels to reduce manufacturing complexity~\cite{eigensatz2010paneling,liu2024reducing}. In rapid prototyping, complex forms can be approximated using simple, reusable bricks, such as LEGO parts~\cite{kim2014survey}, as seen in systems like FaBrickation~\cite{mueller2014fabrickation}. We apply a similar strategy to fashion: our system uses a small set of modules types that can be flexibly recombined to create a wide range of garment features and silhouettes.}

Finally, fast, error-resistant construction is key to usability and accessibility. Prior work has introduced methods to reduce alignment errors in laser-cut assemblies~\cite{park2022foolproofjoint} and accelerate fabrication with hinged or slot-fit joints~\cite{abdullah2021roadkill, abdullah2022hingecore}. Inspired by these works, we employ bidirectional connectors that allow any aligned module edges to attach, enabling quick and simple garment construction.

\section{Sewing Background \& Terminology}\label{sec:background}
Our modular garment system, \garmod, borrows common concepts from conventional sewing.
A \defn{pattern} is a size-specific blueprint for constructing a desired garment.
It comprises outlines for \defn{panels} cut from fabric sheets and instructions for joining their edges in \defn{seams} or leaving them as \defn{free edges}.
Each pattern also specifies a \defn{seam allowance}—a margin between the panel edge and stitching line that provides stability.

\begin{figure}[ht]
\includegraphics[width=\columnwidth]{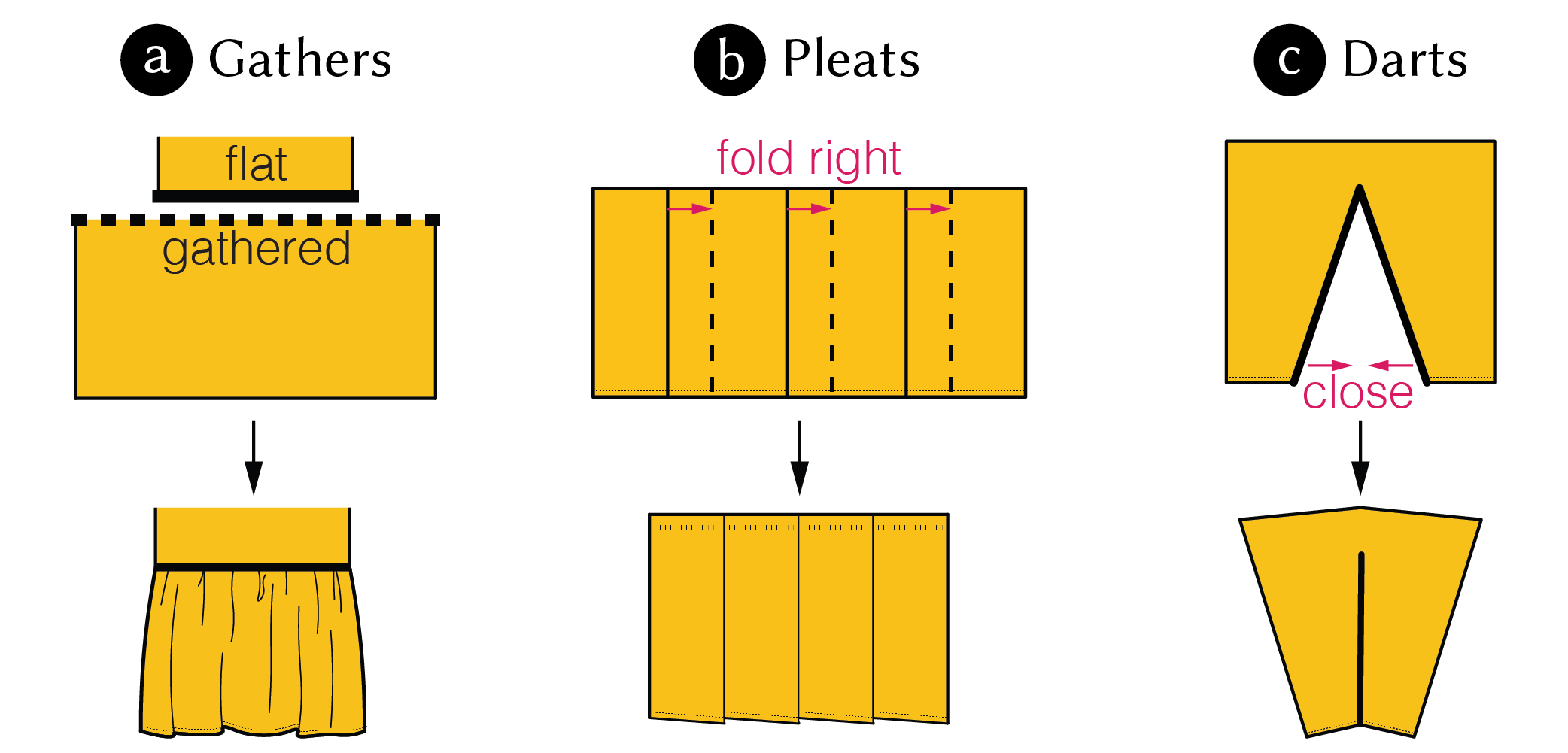}
\caption{\label{fig:features}
Three common fabric-shaping techniques:
(a) gathers join a longer panel edge to a shorter one via soft, evenly spaced folds to add fullness to garments;
(b) pleats double fabric back onto itself in structured and secured folds to vary garment volume and style;
and (c) darts remove wedge-shaped sections of fabric to contour garments to the body.
}
\Description{Three shaping techniques. (A) Gather: attaching a long panel to a shorter one creates soft folds. (B) Pleat: accordion-style folds along a panel. (C) Dart: wedge-shaped removal introduces curvature.}
\vspace{-0.2cm}
\end{figure}

Fabric manipulation techniques beyond basic seaming are used to alter a garment's 3D geometry (\figref{fig:features}).
\defn{Gathers} compresses a longer edge into a shorter one through soft, evenly distributed folds to create fullness. They are commonly applied at waistlines, sleeves, cuffs, or for ruffles.
\defn{Pleats} double and secure fabric over itself in structured folds.
They can span an entire panel, such as in a pleated skirt, or sit at select locations, such as along a trouser waistband.
\defn{Darts} remove triangular wedges of fabric to contour fabric for fit.
They are commonly positioned at the bust, waist, or hips, with their width and height governing the resulting shaping.
In practice, these operations require adjusting the original panel outlines—extending edges to accommodate future gathering, folding, or cutouts—while preserving the garment's intended finished dimensions.

\section{Modular Garment Design}\label{sec:modules}

\edited{\garmod draws inspiration from several creative approaches to promoting garment longevity and reducing textile waste, including mending~\cite{jones2021patching,lakhdhir2024expressive}, zero-waste clothing~\cite{rissanen2016zero, zhang2024wastebanned}, and capsule wardrobes~\cite{mak2017capsule,bardey2022finding, hsiao2018creating, patil2021graph}.
By assembling garments from a small set of rectangular modules, users can repair clothing by replacing only the affected parts, draft patterns with minimal fabric waste, and recombine elements to create new looks. Our \emph{design goals} for modular garments include:}
\aptLtoX{\begin{enumerate}
\item[(G1)] \makeatletter\def\@currentlabel{G1}\makeatother \label{goal:versatile}
\emph{Versatility:}
Support diverse garment shapes and styles with a small set of module types and sizes.
\item[(G2)]\makeatletter\def\@currentlabel{G2}\makeatother\label{goal:reusable}
\emph{Reconfigurability:}
Enable easy disassembly and reconfiguration to adapt to changing bodies and contexts.
\item[(G3)]\makeatletter\def\@currentlabel{G3}\makeatother\label{goal:comfortable}
\emph{Wearability:} Ensure garments fit properly, remain secure, and retain the appearance of conventional clothing.
\end{enumerate}}
{
\begin{goals}
\item\label{goal:versatile}
\emph{Versatility:}
Support diverse garment shapes and styles with a small set of module types and sizes.
\item\label{goal:reusable}
\emph{Reconfigurability:}
Enable easy disassembly and reconfiguration to adapt to changing bodies and contexts.
\item\label{goal:comfortable}
\emph{Wearability:} Ensure garments fit properly, remain secure, and retain the appearance of conventional clothing.
\end{goals}}

Our approach to modular garment design differs from conventional sewing-based construction in two key ways: first, panels are assembled from discrete, reusable modules rather than cut from a single-use continuous sheet; second, seam edges are joined using specialized interfaces, which are carefully arranged snaps or fasteners, that enable reversible, no-sew assembly \ref{goal:reusable}. The remainder of this section introduces these modules and interfaces, outlines the operations they support, and characterizes the resulting design space of modular garments.

\subsection{Module Definitions}
To support a diverse range of garments~\ref{goal:versatile}, we introduce three module types: foundation, pleat, and dart, which enable advanced shaping capabilities similar to those for conventional clothing (\secref{sec:background}). Each module is sized as an integer multiple of a user-defined base unit, $\Delta$, which defaults to 8~cm.

\subsubsection{Foundation Modules}
A \defn{foundation module} is a square building block with dimensions $\D \times \D$ (\figref{fig:foundation}) or an integer multiple thereof, $m\D \times m\D$, where $m \in \mathbb{N}$.
Foundation modules alone are sufficient to construct complete garments, as demonstrated in \figref{fig:system_overview}.

\setlength{\columnsep}{8pt}
\begin{wrapfigure}{r}{0.36\linewidth}
  \vspace{-23pt}
  \begin{center}
    \includegraphics[width=\linewidth]{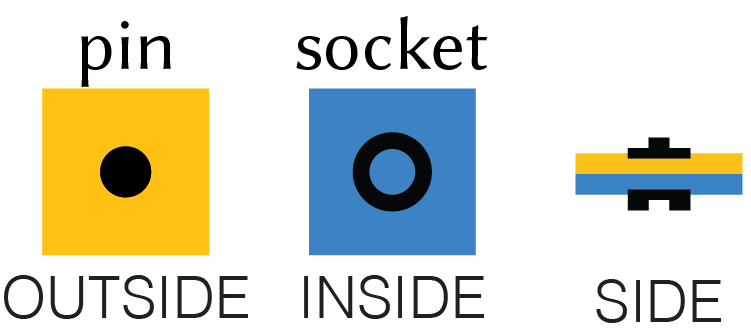}
  \end{center}
  \vspace{-16pt}
\end{wrapfigure}
\subsubsection{Seam Interface}\label{sec:seam-interface}
\begin{figure}[h]
\includegraphics[width=\columnwidth]{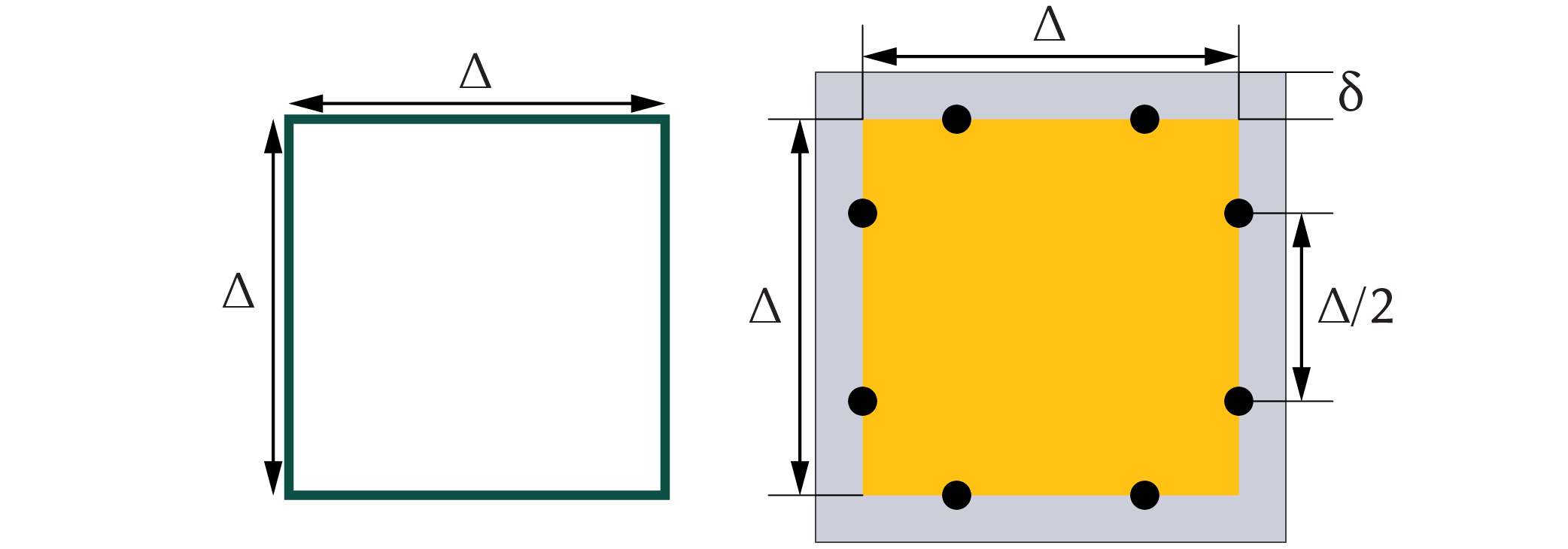}
\caption{\label{fig:foundation}
A $\D \times \D$ foundation module (left) and one possible implementation (right), featuring two connection points per edge and a seam allowance of $\delta$.}
\Description{Foundation module schematic. Left: square module of size Δ × Δ. Right: two snaps per edge, with seam allowance shown in gray.}
\end{figure}

\edited{Each module edge has a \defn{seam interface}: an array of double-sided snaps (inset) or fasteners (\figref{fig:interface-fabrication}), centered along the edge and evenly spaced at intervals of $\Delta/(2k)$, where $k$ is a positive integer that ensures an even number of connection points and modulates their density (e.g., $k=1$ in \figref{fig:foundation}).
}
Similar to conventional seam allowances (\secref{sec:background}), each module has a $\delta$ margin that creates an overlap between modules, preventing gaps when the garment is under tension.

\begin{figure}[h]
\includegraphics[width=\columnwidth]{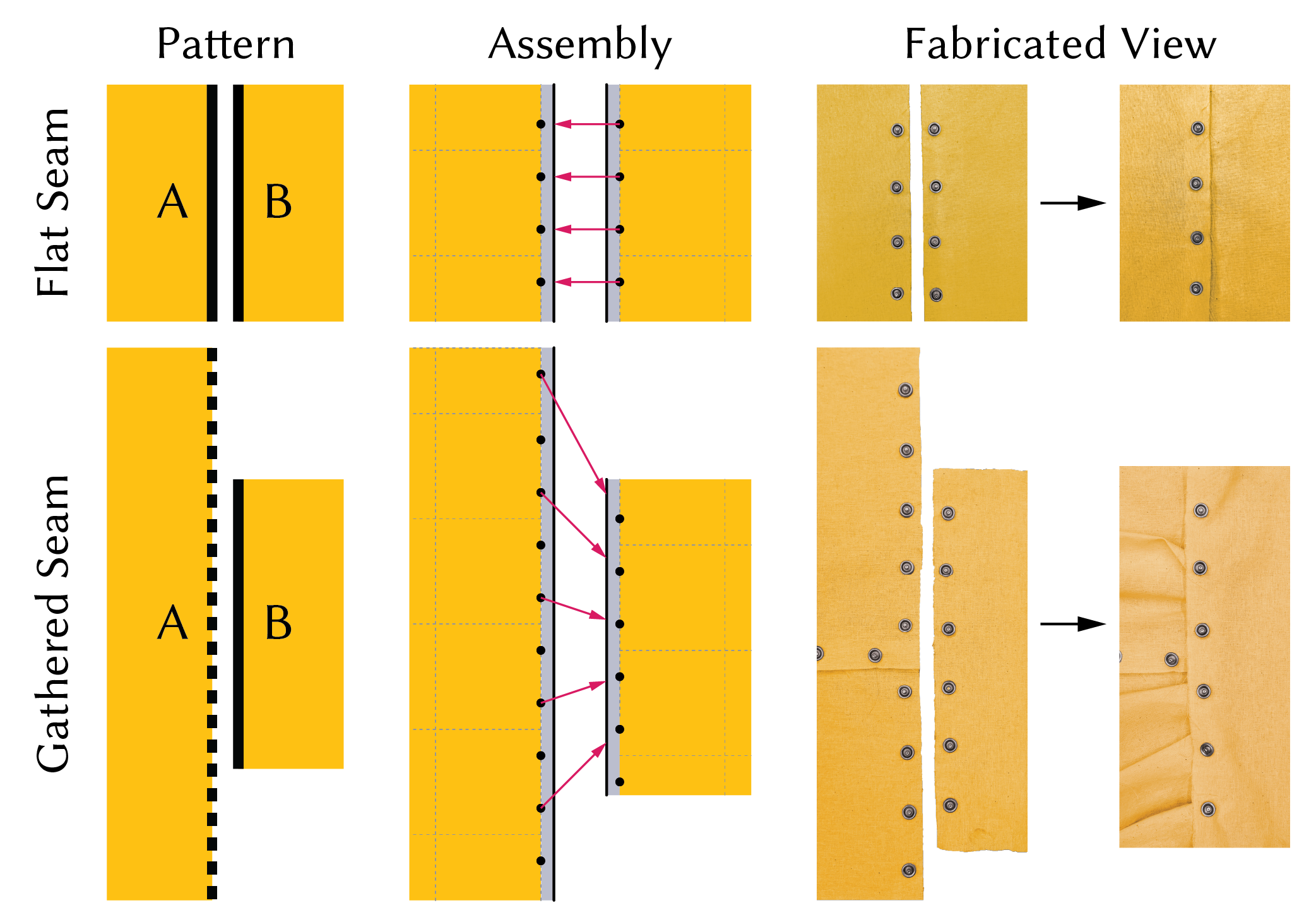}
\caption{\label{fig:seam-types} Our seam interface supports two types of seams. A \emph{flat seam} joins two edges of equal length by aligning fasteners in a one-to-one correspondence. A \emph{gathered seam} connects a longer edge (dashed) to a shorter edge by pairing two fasteners on the longer side with one on the shorter. Because the interface is bidirectional, either panel may lie on top. In the fabricated examples, panel A lies atop B for the flat seam, while B lies atop A for the gathered seam.
}
\Description{Seam interface types. Top: flat seam with one-to-one fastener alignment. Bottom: gathered seam pairing two fasteners on the longer edge to one on the shorter.
}
\end{figure}

The \emph{discrete}, or \emph{point-to-point}, nature of our seam interface enables two types of seams (\figref{fig:seam-types}).
A \defn{flat seam} joins edges of equal length by matching fasteners one-to-one.
A \defn{gathered seam} joins edges in a 2:1 ratio: the longer edge is gathered so that every other fastener aligns with one on the shorter edge.
A seam may be \emph{partially} gathered if it combines both flat and gathered lengths, as seen in \figref{fig:operation-duplicate-column} and \figref{fig:patchwork}.

\subsubsection{Pleat Modules}
\begin{figure}[h]
\includegraphics[width=\columnwidth]{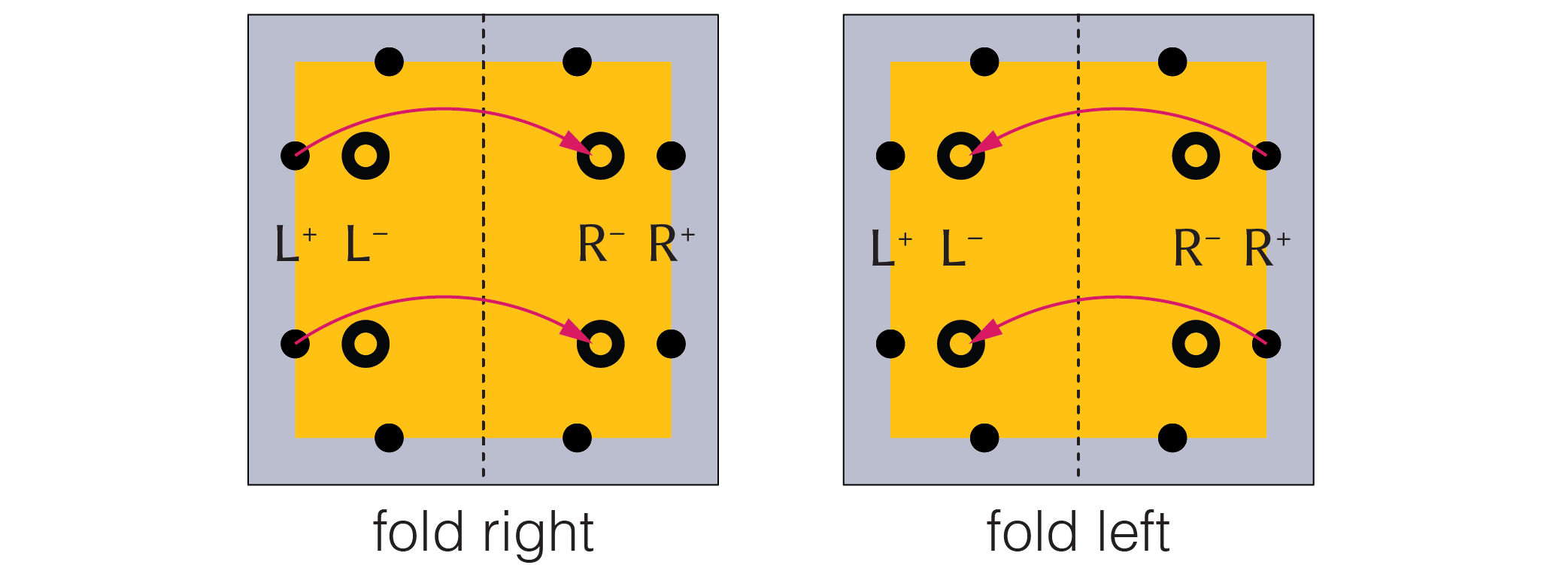}
\caption{A pleat module can be folded in two directions by connecting seam interface pins to pleat interface sockets: (1) right by connecting left pins to right sockets ($L^{+} \rightarrow R^{-}$), or (2) left by connecting right pins to left sockets ($R^{+} \rightarrow L^{-}$).}\label{fig:pleat}
\Description{Pleat module with directional folds. Left: connecting left pins to right sockets folds rightward. Right: connecting right pins to left folds leftward.
}
\end{figure}

\begin{figure}[h]
\includegraphics[width=\columnwidth]{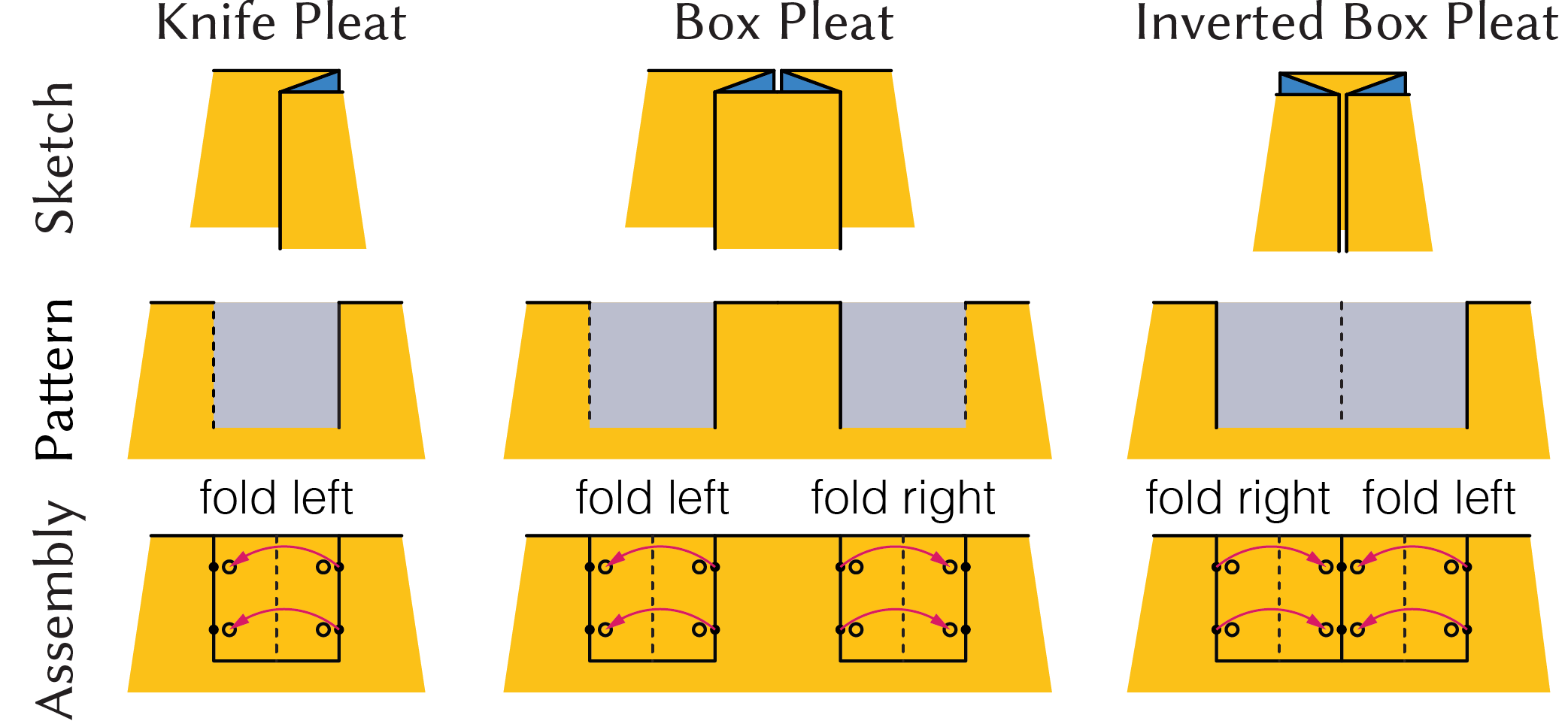}
\caption{\label{fig:pleats}Arrangements of pleat modules to achieve knife, box, and inverted box pleats.}
\Description{Pleat arrangements. (A) Knife pleats. (B) Box pleat. (C) Inverted box pleat.
}
\end{figure}

A \defn{pleat module} is identical to the $\D \times \D$ foundation module except for an additional \defn{pleat interface} that supports a directional fold. The interface comprises four sockets located on the front face of the module, immediately adjacent to the pins of the left and right seam interfaces (\figref{fig:pleat}).
\figref{fig:pleats} illustrates pleating effects produced by different arrangements of pleat modules.

\subsubsection{Dart Modules}
\begin{figure}[h]
\centering
\includegraphics[width=\columnwidth]{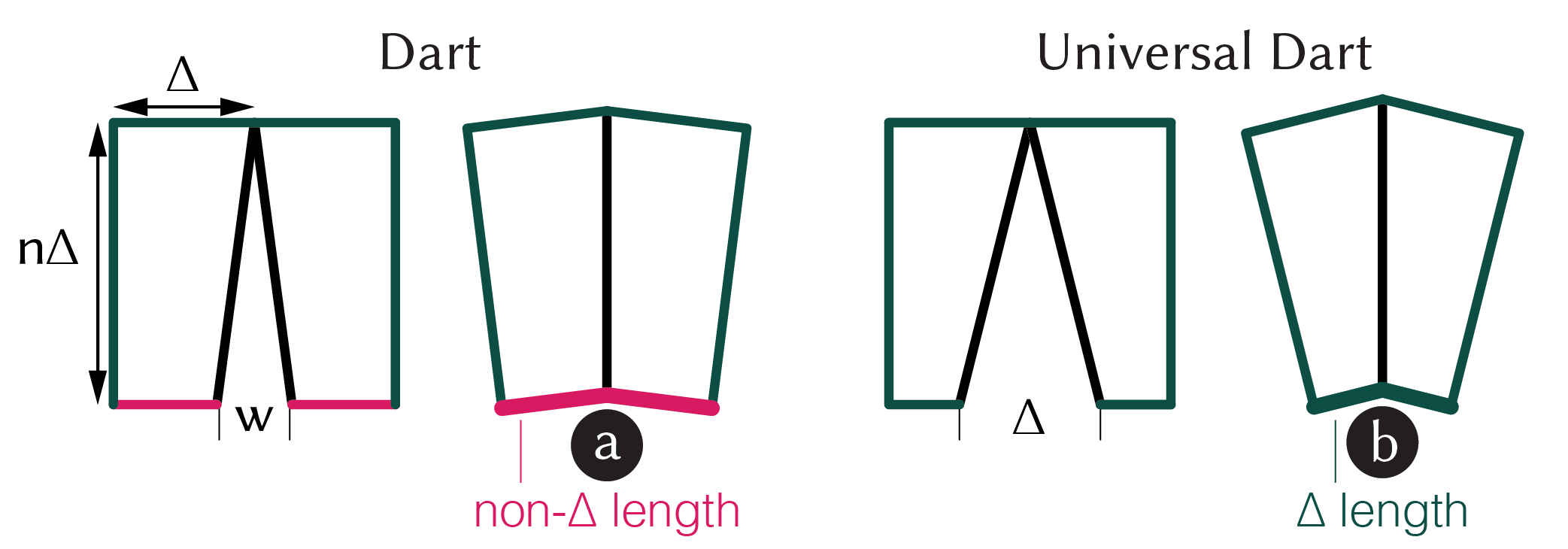}
\caption{\label{fig:dart}Left: A dart module with width \(w\) and height \(n\Delta = 2\D\). Right: A \emph{universal} dart module (\(w = \Delta\)) ensures that all assembled edge lengths remain integer multiples of $\D$; edge (b), unlike (a), has length $\Delta$, and so it may be part of a seam.}
\Description{Standard vs. universal dart modules. Left: standard dart causes a fractional edge. Right: universal dart yields an integral folded length for grid and seam alignment.
}
\end{figure}

A \defn{dart module} comprises two mirrored right-angle trapezoids whose slanted legs form a triangular gap; this gap is folded closed during assembly, creating 3D shaping.
Each trapezoid is parameterized by a height parameter \(h = n\Delta\), where \(n \in \mathbb{N}\), a wide base of length \(\Delta\), and a narrow base of length \(\Delta - \tfrac{w}{2}\) for some width parameter \(w\), subject to \(w \le 2\Delta\). Once assembled, the dart module's narrow end measures \(2\Delta - w\) (\figref{fig:dart}a). If \(w = \Delta\), this length is \(\Delta\) and is therefore compatible with the $\Delta$-aligned dimensions of foundation and pleat modules, allowing it to participate freely in seams. We call this the \defn{universal} case. Otherwise, for any other value of \(w\), the module can either form a triangle dart along a free edge or pair with another dart module of the same width to create a diamond-shaped dart.
Attaching a non-universal dart module to a $\Delta$-aligned edge would result in a seam mismatch of length \(\pm w\) and is therefore not permitted.

\begin{figure*}[t]
\centering
\includegraphics[width=\textwidth]{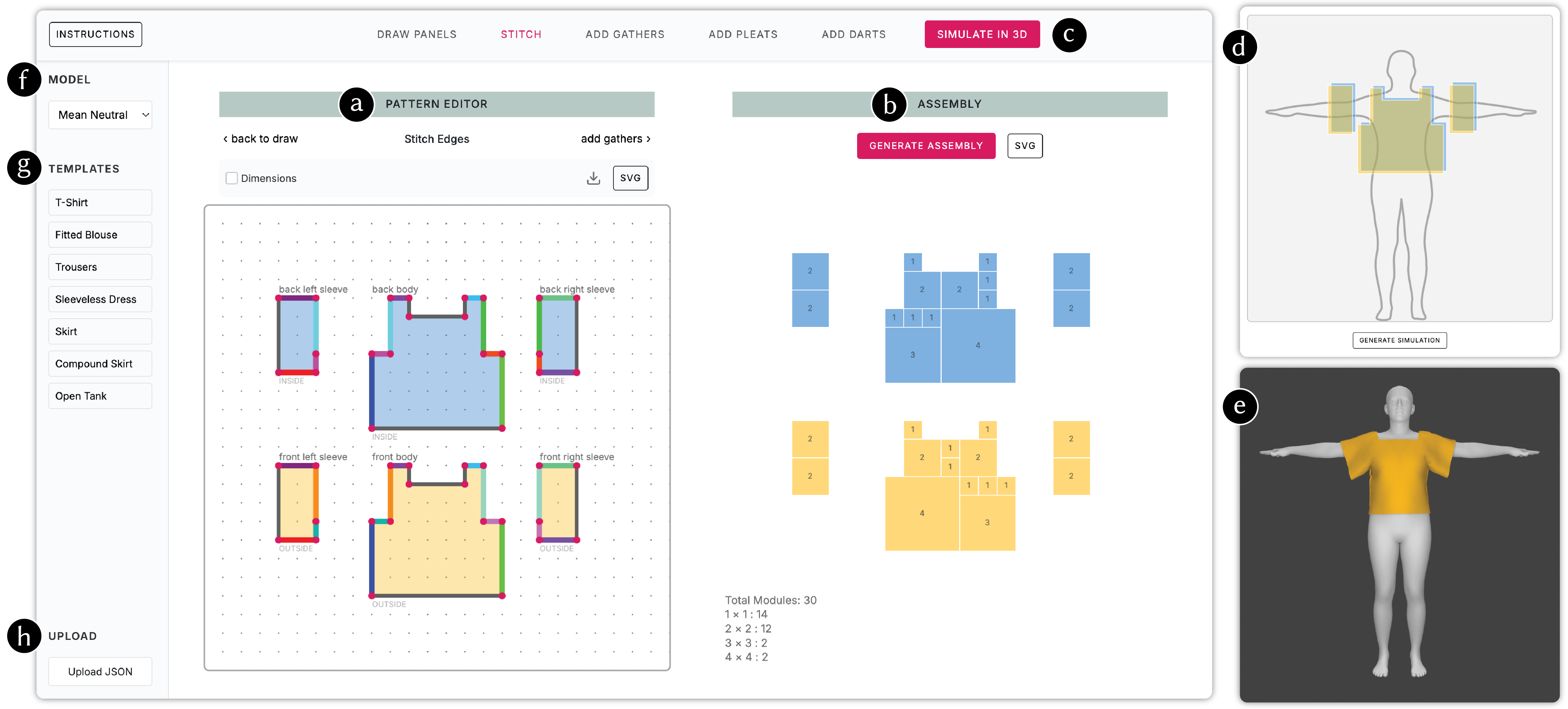}
\caption{\label{fig:system_overview}
\garmod's interface consists of three main views:
(a) the \textit{Pattern Editor} for drawing and stitching garment panels and adding shaping features;
(b) the \textit{Assembly View} for decomposing the pattern into a modular assembly;
and (c) the \textit{Simulation View} for aligning (d) and previewing (e) the garment on a selected body model (f).
Users can begin their design process by drawing a pattern from scratch (a), selecting from existing templates (g), or uploading a previously created pattern (h).
}
\Description{Refashion Studio interface. (A) Pattern Editor. (B) Assembly View. (C) 3D Simulation. Other features: (D) pattern alignment, (E) garment draping, (F) avatar selection, (G) garment templates, (H) pattern import.}
\end{figure*}

\subsection{Modular Patterns}\label{sec:design_space}

Similar to traditional garments, our modular garments are built by joining collections of 2D panels. However, in our approach, each panel is first assembled from a set of foundation, pleat, and dart modules. To ensure that a design can be realized using only these modules, we enforce two constraints on any \defn{modular pattern}:

\aptLtoX{
\begin{enumerate}\label{modular-constraints}
\item[(C1)]\makeatletter\def\@currentlabel{C1}\makeatother \label{constraint:geometry}
Panels, including any pleats or darts, must align with a regular grid of cell size~$\D$.
\item[(C2)]\makeatletter\def\@currentlabel{C2}\makeatother \label{constraint:matching}
Seams may only join module edges whose lengths differ by at most a factor of two (\secref{sec:seam-interface}).
\end{enumerate}}{
\begin{constraints}\label{modular-constraints}
\item\label{constraint:geometry}
Panels, including any pleats or darts, must align with a regular grid of cell size~$\D$.
\item\label{constraint:matching}
Seams may only join module edges whose lengths differ by at most a factor of two (\secref{sec:seam-interface}).
\end{constraints}}

An \defn{assembly} is a decomposition of a modular pattern into modules (e.g., \figref{fig:system_overview}b) from which the pattern can be constructed. An assembly is considered efficient when it uses as few modules as possible, serving as a proxy for assembly effort.
In other words, our goal is to divide the garment panels that comprise the pattern into grid-aligned regions, possibly with holes, with the fewest available modules. We formalize and solve this problem in \secref{sec:assembly_formulation}.

\section{\garmod~UI Overview}\label{sec:system}

We present \garmod~Studio, a digital tool for designing modular garment patterns.
The UI comprises three main views:
the \emph{Pattern Editor View} (\figref{fig:system_overview}a) for creating and editing patterns,
the \emph{Assembly View} (\figref{fig:system_overview}b) for generating pattern assembly instructions,
and the \emph{Simulation View} (\figref{fig:system_overview}c) for previewing garments in 3D.
\edited{The tool assumes no prior pattern-making experience and includes comprehensive user guides with video walkthroughs for pattern creation and shaping.}
\figref{fig:walkthrough} illustrates the step-by-step creation of a compound skirt. In Phase 1, the base pattern is drawn and stitched. In Phase 2, gathers, pleats, and darts are applied to shape the garment (see \appref{app:feature-casework} for full feature manipulation details).

\subsection{Pattern Editor View}

Users begin their design process by either
selecting a pattern from the tool's built-in library (\figref{fig:system_overview}g),
uploading a previously saved pattern (\figref{fig:system_overview}h),
or creating a new pattern from scratch (\figref{fig:system_overview}a).

When creating a new pattern, users first draw garment panels as non-self-intersecting loops of horizontal and vertical segments aligned to a
$\D$-unit grid~\ref{constraint:geometry}. The system automatically validates each panel's geometry. Users can rename panels, drag them into place, or flip their fabric orientation. These basic properties are locked once the user enters the stitching phase, where they select pairs of panel edges with matching flat lengths to form seams. If a single edge must be divided—for example, to participate in multiple seams or to leave part of it free—users can insert break points to split it into segments.

After creating a base pattern (Phase 1), users can add shaping features in the order of gathers, then pleats, and finally darts (Phase 2). This orders the features from the most to the least disruptive to a panel's dimensions: a gather doubles an edge's length, a pleat may introduce and then fold away a single unit of length, and a dart alters only local geometries.

\begin{figure*}[h]
\includegraphics[width=\textwidth]{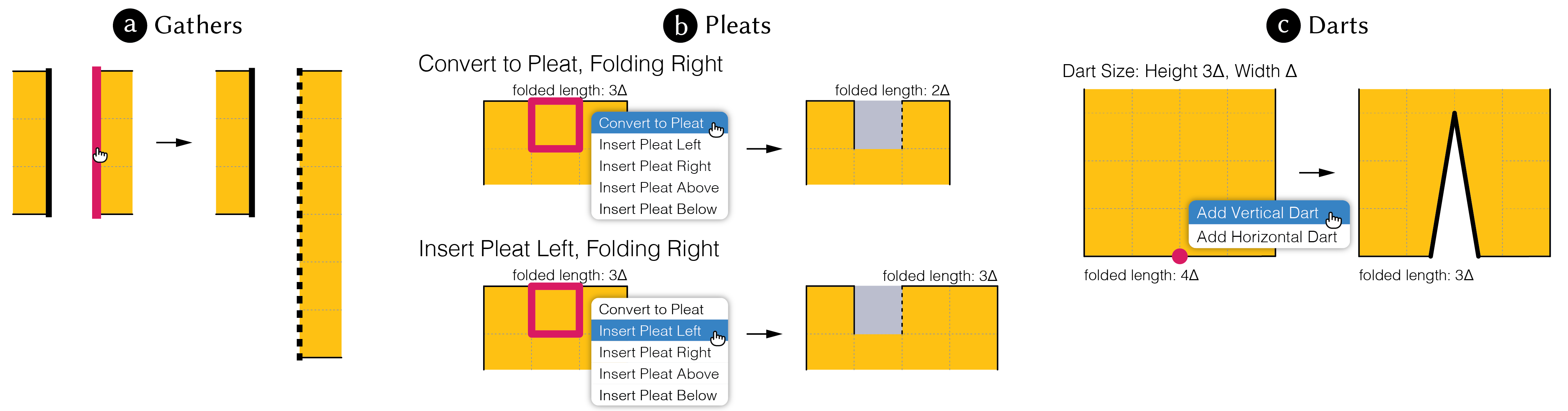}
\caption{\label{fig:ui-feature-operations}
\edited{In the Pattern Editor View,
(a) gathering an edge at a seam doubles its length.
(b) Converting an existing unit into a pleat module reduces the edge's folded length, whereas inserting a new pleat preserves the folded length but increases the flat length.
(c) Adding a dart replaces the local geometry with the specified dart module. More complex examples with cascading effects on seam relationships appear in \appref{app:feature-casework}.}}
\Description{Pattern Editor feature interactions. (A) Gathering doubles seam length. (B) Pleat conversion shortens folded length; insertion adds flat length. (C) Dart insertion alters local geometry only.
}
\end{figure*}

\subsubsection{Gathers}
To gather an edge in a seam (\figref{fig:seam-types}), the system doubles its length (\figref{fig:ui-feature-operations}a). Gathering may affect adjacent seams; \garmod~Studio automatically propagates these adjustments or, if a gather is not feasible at that location, alerts the user to the conflict (\appref{app:gather_details}).

\subsubsection{Pleats}
Users have two options when adding pleats: they can either convert an existing panel unit into a pleat, which reduces the panel edge's folded length, or insert a new pleat unit to maintain the current folded length (\figref{fig:ui-feature-operations}b). Using the menu above the pattern editor, users specify the pleat's fold direction (up, down, left, or right). Like gathers, pleats may require adjustments to existing seams (\appref{sec:pleat_details}). \garmod~Studio validates pleat placement and alerts users if a location cannot support the desired pleat.

\subsubsection{Darts}
\begin{figure*}[h]
\includegraphics[width=\textwidth]{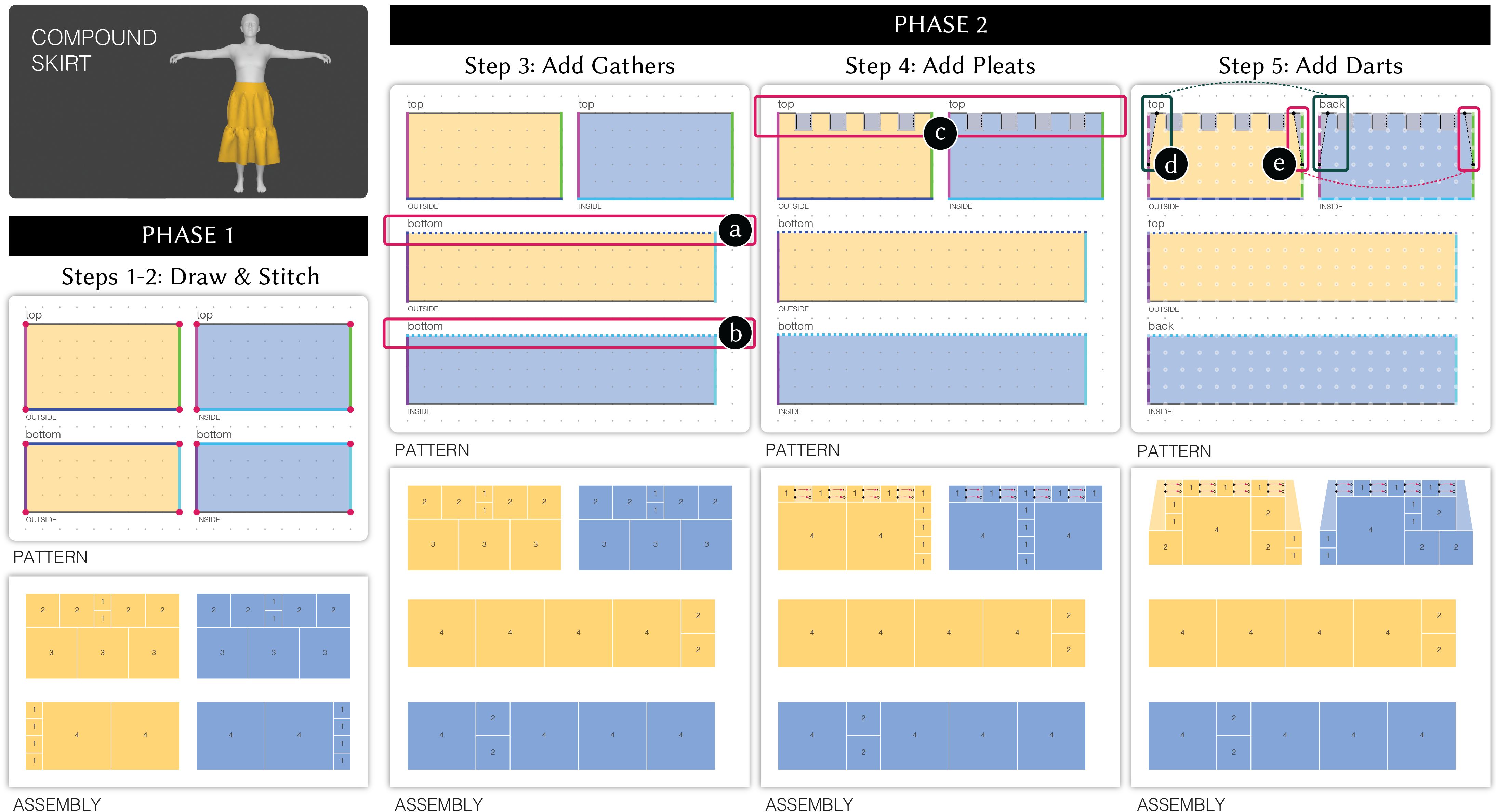}
\caption{\label{fig:walkthrough} To draw a two-layer compound skirt in \garmod~Studio, the user starts by drawing and stitching panels to form the base pattern as shown in Phase 1 (Steps 1-2). In Phase 2, they manipulate its shape by adding features. First (Step 3), they gather edges (a) and (b) to add volume to the second layer of the skirt. Next (Step 4), they convert every other unit along its waistline (c) into right-folding pleats to cinch the waist. Finally (Step 5), they add darts (d) and (e), which align across the side seams, to refine the waist-to-hip shaping. For each step, we display both the resulting pattern (top) and its corresponding assembly (bottom). The simulation is in the upper left.}
\Description{Compound skirt construction. Step 1: panel drawing and stitching. Step 2: feature addition—gathers for volume, pleats for waist cinching, darts for contouring. Insets show 3D simulations.}
\end{figure*}

To create a dart, users click any grid point within a panel and choose between a horizontal or vertical orientation (\figref{fig:ui-feature-operations}c).
A dropdown menu above the editor allows users to specify the dart's height and width.
\garmod~Studio validates the placement and adds the dart if possible (\appref{sec:dart_details}), as shown in \figref{fig:ui-feature-operations}c and \figref{fig:walkthrough}, or explains why the placement is invalid.

\subsection{Assembly View}
If the user has specified a set of available input modules, the Assembly View automatically decomposes the design into a pattern that only uses these modules.
If the user does not specify existing modules, the system assumes the user will cut the required modules.
Given a decomposition, users can review the suggested modules and download an SVG cutting guide to make what is needed.

\subsection{Simulation View}
The Simulation View enables interactive 3D visualization of garment designs (\figref{fig:system_overview}c). \garmod~Studio provides a library of body models across multiple sizes~\cite{pavlakos2019expressive} (\figref{fig:system_overview}f) and allows users to import custom T-pose OBJ meshes for personalized fitting. To initiate a simulation, users drag and align each pattern panel to the model's silhouette (\figref{fig:system_overview}d). \edited{Once the simulation is running, users can rotate and zoom the 3D scene to inspect the garment from any view.}

\section{Implementation}\label{sec:implementation}

In this section, we discuss the underlying implementation for representing and decomposing modular sewing patterns. First, we present a grid-based pattern representation that supports modular pattern design under strict panel and seam constraints (\secref{sec:design_space}). Next, we discuss our approach to generating efficient modular garment assemblies by formulating the problem as an integer linear program. For details on garment simulation, please refer to \appref{app:simulation}.

\subsection{Pattern Creation}\label{sec:pattern-representation}

\begin{figure}[h]
\includegraphics[width=\columnwidth]{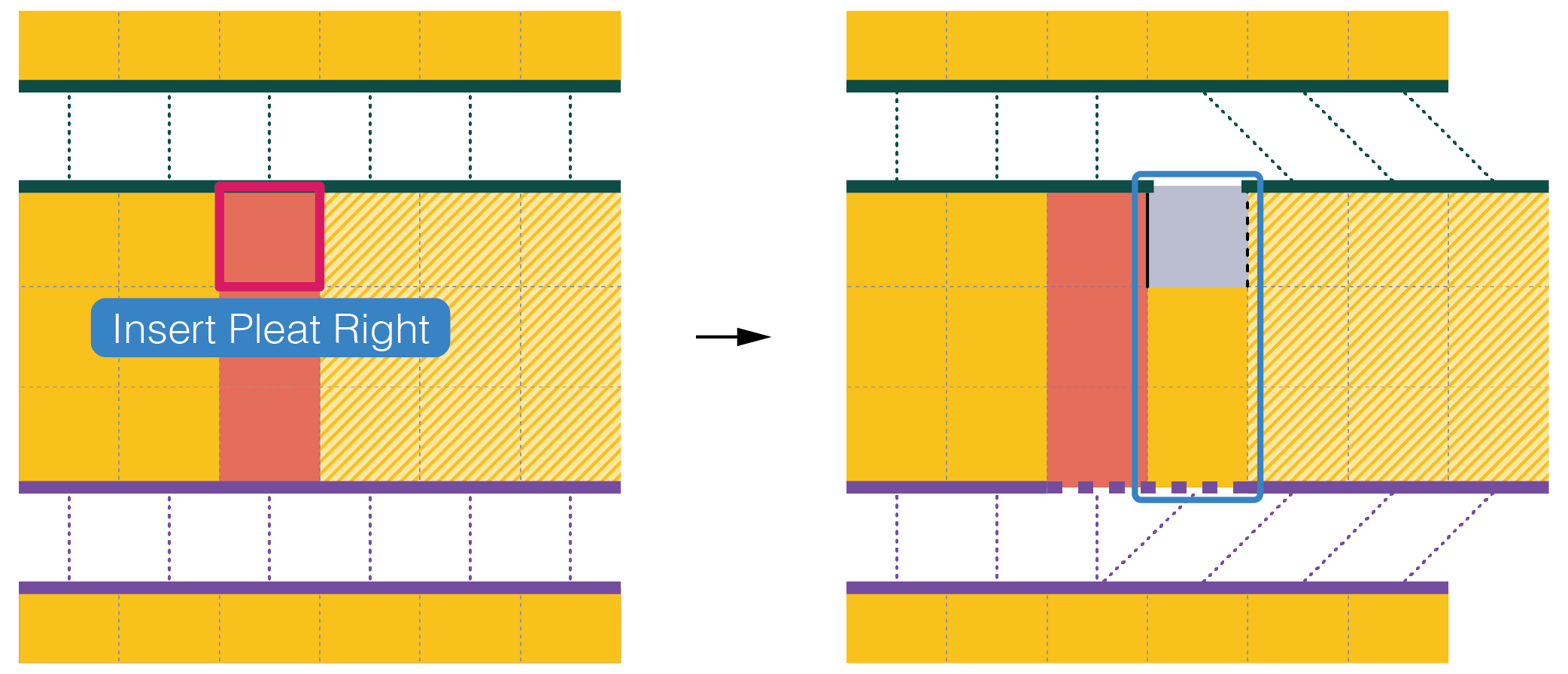}
\caption{\label{fig:operation-duplicate-column}
Inserting a pleat (gray square) to the right of a selected square (red outline) updates the panel geometry.
Left: The system highlights the strip containing the selected square.
Right: Adjacent squares (shaded) shift to make room for a duplicate strip (outlined). This insertion triggers seam rematching and introduces a gather (dashed) along the bottom seam to accommodate the added length.}
\Description{Pattern update upon pleat insertion. Left: selected strip highlighted. Right: adjacent modules shift to insert a duplicate; bottom seam gather compensates for added length.
}
\vspace{-0.3cm}
\end{figure}

A pattern is shown to the user at two levels of granularity:
a coarse level for creating a base pattern (Phase 1)
and a finer-grained level for localized feature (e.g., pleat and dart) manipulation (Phase 2).
In Phase 1, each panel is defined by a list of edges and a binary orientation for the face of the module visible in the pattern. When users begin feature edits in Phase 2 (\figref{fig:walkthrough}), we switch to a $\D$-grid view to enforce the two key constraints from \secref{sec:design_space}: panels must be $\D$-grid-aligned polygons~\ref{constraint:geometry}, and seams may differ in length by at most a factor of two~\ref{constraint:matching}.
Each panel is decomposed into a collection of $\D$-sized square units, and its edges into $\D$-length segments, each tracking local connectivity and, if relevant, seam pairings (e.g., dotted lines in \figref{fig:operation-duplicate-column}).

\paragraph{Panel Edits via Strip Operations}
Working with gathers and pleats often requires modifying panel geometry to accommodate the additional material (\figref{fig:ui-feature-operations}). To preserve orthogonality and $\D$-aligned dimensions, we perform these operations using \defn{strips}—maximal sets of connected squares along a single row or column.

When a panel is expanded or contracted through the addition or removal of gathers or pleats, we insert or delete entire strips, ensuring that all edges remain axis-aligned and multiples of $\D$. For example, inserting a pleat duplicates the strip containing the selected square and places the copy adjacent to the original (\figref{fig:operation-duplicate-column}). Similarly, gathering an edge duplicates all incident strips and so doubles its length.

The strip insertion process begins with a flood-fill operation from the strip's boundary to identify affected squares. We then shift these squares by one $\D$ unit, creating space for the duplicated strip, and update adjacency pointers to preserve the panel's topology. This localized approach prevents disconnections that might occur with full-column shifts, particularly in panels with cavities. While strip deletion follows a similar process in reverse, we prohibit deletions that would disrupt existing seams or features.

Because inserting or deleting strips changes edge lengths, any affected seams must be rematched. If a proposed strip operation would violate seam matching constraints—for example, by attempting to fold away a portion of an edge whose match is already fully gathered—the operation is prohibited unless additional edits first resolve the conflict. Please refer to \appref{app:feature-casework} for a more detailed discussion of these operations.

\paragraph{Seam Re-Matching Upon Panel Edits}
At the start of feature edits, we mark each edge segment in a seam as active and establish a one-to-one, \emph{planar} (non-crossing) correspondence between opposing segments—ensuring that the seam does not ``twist'' in our representation. These matchings evolve as the user makes edits: an active segment may gain a second match through gathering, lose all matches by becoming inactive in a fold, or retain a single match. If every segment of an edge already has two matches, we prohibit any feature operation that would shorten it or lengthen its counterpart. More generally, feature operations are allowed only if they preserve one or two matches per active segment across all affected edges. When updates are required, we adjust matchings as locally as possible while preserving planarity. The following algorithm enforces these constraints.

\begin{figure}[h]
\includegraphics[width=\columnwidth]{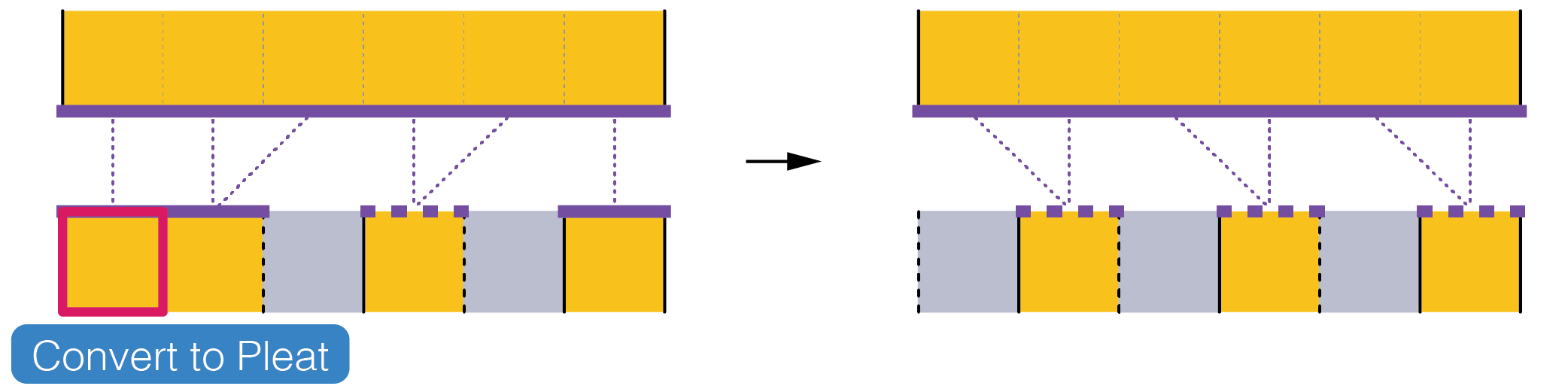}
\caption{\label{fig:operation-rematch} Pleating along a partially gathered seam may trigger cascading rematches to preserve seam constraints.}
\Description{Seam rematching after pleat conversion along a partially gathered seam, with cascading adjustments preserving matching invariants.}
\end{figure}

If inserting a new segment along an edge (e.g., via strip insertion; \figref{fig:operation-duplicate-column}) results in an equal number of active segments on both edges, we establish new one-to-one correspondences directly. Otherwise, we create a sequence of slots—up to two per active segment—along the edge and shift existing matches within these slots, based on the insertion location and seam orientation, to accommodate the new segment. These shifts may cascade to preserve matching invariants. Similarly, when deactivating a segment during folding or dart insertion, we redistribute its matches to neighboring segments, which can also trigger cascading adjustments (\figref{fig:operation-rematch}).

\subsection{Assembly Formulation}\label{sec:assembly_formulation}

\edited{
To construct a modular pattern, we select a minimal set of modules that assemble into the desired layout.
Once the user has set the pleat and dart locations, the remaining task is to fill the pattern with foundation modules.
Our seam interface (\secref{sec:seam-interface}) is designed to be fully \emph{bidirectional}, allowing any two module edges to connect.
As a result, all foundation modules of the same size are \emph{interchangeable}. This reduces the task to square covering, which we formalize below as an integer linear program (ILP), using $\Delta = 1$ for illustration.
}

\paragraph{Problem Formulation}
Let \(\mathcal{S}\) be a set of foundation modules: squares of specified positive integer side lengths, and let \(\mathcal{P} = \{P_1, \dots, P_K\}\) denote the remaining garment pattern as a collection of grid-aligned polygons, possibly with holes.
The objective is to cover \(\mathcal{P}\) with a non-overlapping subset \(U \subseteq \mathcal{S}\) that minimizes \(|U|\).

This partitioning problem is generally intractable;
even packing a maximal number of \(2\times2\) arrays into a rectilinear grid is NP-complete~\cite{fowler1981optimal,hochbaum1985approximation}.
However, we formulate our task as an ILP and solve practical instances both optimally and efficiently. We empirically validate this performance across a range of pattern sizes and shapes in \appref{apx:decomp_eval}.

\paragraph{ILP Formulation}

Let $n_{\ell}$ denote the number of available squares of side length $\ell$ in the collection \(\mathcal{S}\). A \emph{candidate} for polygon \(P_k\) is specified by a side length $\ell$ (with $n_{\ell}>0$) and a lower-left coordinate \((i,j)\in\Z^2\). The corresponding square is defined as
\[
S(\ell,i,j) = [i,\, i+\ell] \times [j,\, j+\ell].
\]
It must lie fully in \(P_k\). Let \(\mathcal{C}_k\) denote the set of all such candidates.

We introduce a binary variable per candidate \((\ell,i,j) \in \mathcal{C}_k\):
\[
x^{(k)}_{\ell,i,j} = \begin{cases}
1, & \text{if } S(\ell,i,j) \text{ is selected for } P_k,\\[1mm]
0, & \text{otherwise.}
\end{cases}
\]
These variables are defined by side length~$\ell$, not individual squares, to reduce variable count and improve ILP efficiency.

Let \(G_k\) be the set of unit grid cells contained in \(P_k\). For each cell \((u,v)\in G_k\) and candidate \((\ell,i,j) \in \mathcal{C}_k\), define a binary variable
\[
A_{\ell,i,j}^{(u,v)} = \begin{cases}
1, & \text{if } [u,\, u+1] \times [v,\, v+1] \subseteq S(\ell,i,j),\\[1mm]
0, & \text{otherwise.}
\end{cases}
\]
This variable encodes whether the unit square in $P_k$ with bottom-left corner at $(u, v)$ is covered by candidate $(\ell, i, j)$.

The ILP formulation is as follows:
\begin{align*}
\min \quad & \sum_{k=1}^{K} \sum_{(\ell,i,j) \in \mathcal{C}_k} x^{(k)}_{\ell,i,j} \notag \\
\text{subject to}
& \sum_{(\ell,i,j) \in \mathcal{C}_k} A_{\ell,i,j}^{(u,v)}\, x^{(k)}_{\ell,i,j} = 1,
    && \forall\, (u,v) \in G_k,\; k = 1, \dots, K, \\
& \sum_{k=1}^{K} \sum_{\substack{(i,j) \\ (\ell,i,j) \in \mathcal{C}_k}} x^{(k)}_{\ell,i,j} \leq n_{\ell},
    && \forall\, \ell \text{ with } n_{\ell} > 0, \\
& x^{(k)}_{\ell,i,j} \in \{0,1\},
    && \forall\, k = 1, \dots, K,\; (\ell,i,j) \in \mathcal{C}_k.
\end{align*}

The objective to be minimized is the number of squares used in the covering. The first set of constraints ensures that every unit cell is covered exactly once. The second set enforces that the number of selected squares of each side length does not exceed the available supply. In this formulation, if \(x^{(k)}_{\ell,i,j}=1\), then \(S(\ell, i, j)\) is selected for \(P_k\); that is, a square of side length \(\ell\) is placed in \(P_k\) with its lower-left corner at \((i, j)\).

We solve the ILP using PuLP~\cite{Mitchell2011PuLP} with the CBC solver~\cite{Forrest2005Cbc}. Most instances solve in under a second on a MacBook Air (M2, 2023) with 8~GB of RAM. A detailed evaluation is provided in \appref{apx:decomp_eval}.

\section{Results}\label{sec:results}

With \garmod~we designed and fabricated a variety of modular garments to demonstrate a range of potential use cases (\secref{sec:ourdesigns}). We also invited additional designers to create their own modular garments (\secref{sec:userdesigns}).

\subsection{Module Fabrication}\label{sec:fabrication}
\edited{In developing our modular garments, we explored a range of different textiles and connection mechanisms. For clarity, we use ``fabrication'' to refer to cutting modules and installing their interfaces, and ``assembly'' to refer to joining modules into panels and garments. All modules can be fabricated without sewing.}

\subsubsection{Textiles}
We used cotton muslin (136~gsm), an inexpensive prototyping fabric, for most of our modules. \edited{\figref{fig:teaser} features Kona cotton (147~gsm), a higher-quality, more tightly woven option. Figs.~\ref{fig:repurpose} and~\ref{fig:patchwork} employ much heavier, thicker, and more varied materials: denim from donated jeans and hand-crocheted granny squares.}

\subsubsection{Cutting}
Unless otherwise stated, all modules adhere to a base unit size of $\D=8$~cm with a seam allowance of $\delta=1$~cm. Foundation modules range from $1\D$ to $4\D$ in size, while dart modules are fixed at $\D$ in width and either $2\D$ or $3\D$ in height. We found that this limited set is sufficient to efficiently form a variety of garments with adequate shaping, even across different sizes.

Modules can be cut using scissors or rotary tools, but for speed and precision, we primarily used a standard laser cutter. The regular geometries of the modules streamline cutting and reduce material waste. \edited{Foundation and pleat modules are square, while dart modules consist of trapezoid pairs that pack efficiently as rectangles. As a result, modules can be laid out and cut from fabric sheets with minimal (and sometimes even no) waste (see \appref{app:fabrication}).}

\begin{figure*}[b]
\includegraphics[width=\textwidth]{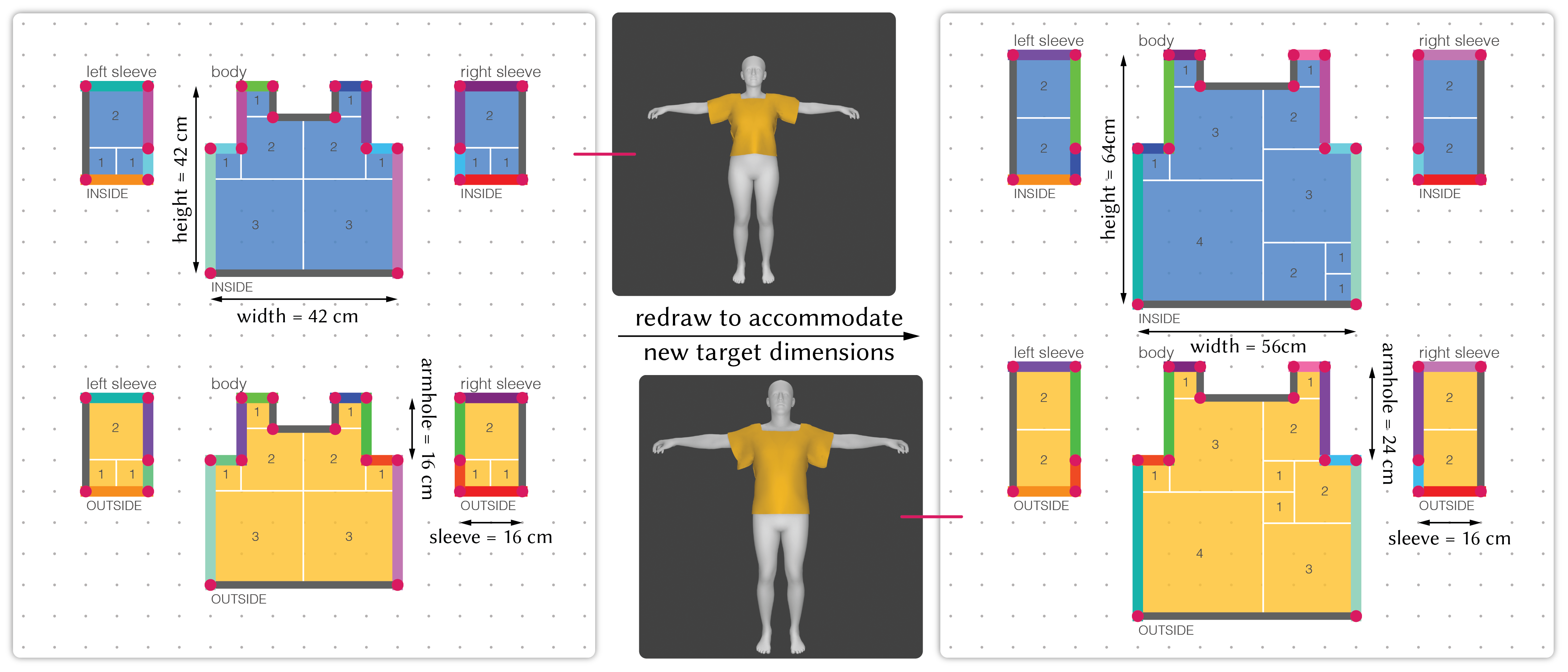}
\caption{\label{fig:resize}
\edited{Given a fixed module set with base unit size $\Delta$, resizing an existing garment pattern (left) involves redrawing it to match new target dimensions (right), rounded to increments of $\Delta$.}}
\Description{Grid-based resizing. Original and resized T-shirt patterns shown side by side; dimensions snapped to the Δ grid.}
\end{figure*}

\edited{\subsubsection{Seam Interface}
We explored two implementations of our seam interface: double-sided connectors and fasteners, which have trade-offs across fabrication time, assembly ease, and wearability.}

\begin{figure}[h]
\includegraphics[width=\columnwidth]{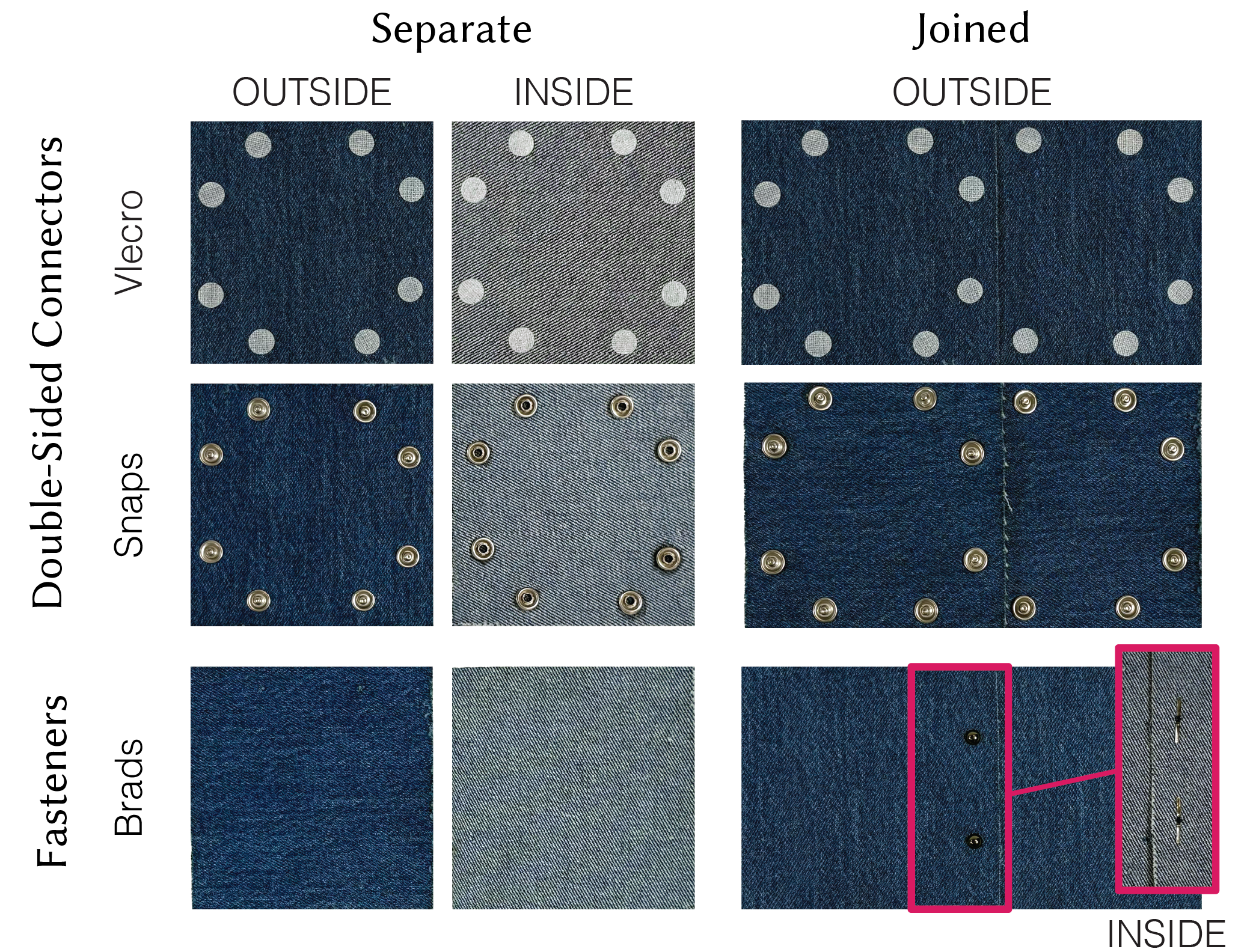}
\caption{\label{fig:interface-fabrication} \edited{Fabricated examples of two seam interface implementations: double-sided connectors using Velcro dots and snap buttons (top), and symmetric fasteners—brads (bottom).}}
\Description{Physical seam-interface prototypes. Top: Velcro dots and snap-button pairs. Bottom: brad fasteners threaded through fabric.}
\end{figure}

\edited{\paragraph{Double-sided Connectors} We tested two connector types (both 9~mm in diameter): adhesive Velcro dots for rapid prototyping and metal snap buttons for added durability. Velcro dots are applied with hooks facing outward and loops inward (\textasciitilde5~sec per installation), and snap buttons are installed by stamping pins onto the module exterior and hot-gluing sockets inside (\textasciitilde1~min per installation) (\figref{fig:interface-fabrication}).
While snaps offer stronger tear resistance, they add bulk; joined seams stack four snaps, creating thick, heavy connections.}

\edited{\paragraph{Fasteners} We tested brads (0.6~cm), which are inserted through module edges and bent open to secure the layers during assembly (\figref{fig:interface-fabrication}). Although assembly takes two to three times longer than with snaps, the resulting garments are lighter, with more secure and less visible connectors (Figs.~\ref{fig:teaser},~\ref{fig:patchwork}–\ref{fig:repurpose}). This method also allows modules to be flipped, as it does not distinguish between inside and outside surfaces, enabling more flexible or even reversible designs.}

\edited{
\subsubsection{Prototype Modules}
We built both lightweight prototypes and more durable garments made from heavier weight materials. The prototypes (e.g., \figref{fig:user-designs}) are assembled from inexpensive muslin modules using Velcro, applied at minimal density—two dots per $\D$-length edge.
Although the prototypes may not be suitable for everyday wear, they can serve as valuable, low-cost building blocks for iterating toward higher-quality garments—modular or otherwise (\secref{sec:prototyping}).
The more polished constructions (e.g., \figref{fig:teaser}) use higher-quality fabrics with fasteners applied at greater density.}

\subsection{Applications of Modular Garments}\label{sec:ourdesigns}

\garmod~leverages key advantages of modular design, including ease of resizing, restyling, and repair. We demonstrate how these adjustments can be efficiently supported through our approach.

\subsubsection{Resizing}\label{sec:resize}
Users can resize a garment by redrawing its pattern to match new target dimensions (\figref{fig:resize}) which would then be assembled from a new combination of modules.
Fine-tuned fit adjustments can be made by adding gathers, pleats, and darts (\figref{fig:walkthrough}).

\subsubsection{Interchanging Parts}\label{sec:interchange}
Our modular system allows users to exchange garment panels or parts within or across designs. For example, panels in maternity wear can be swapped for larger ones to accommodate a growing body without requiring an entirely new garment (\figref{fig:user-designs}-P2), and sleeves from different tops can be mixed and matched—expanding styling possibilities through combinations, even with a fixed set of components.

\subsubsection{Style Adjustments}\label{sec:extend}
\begin{figure}[t]
\includegraphics[width=\columnwidth]{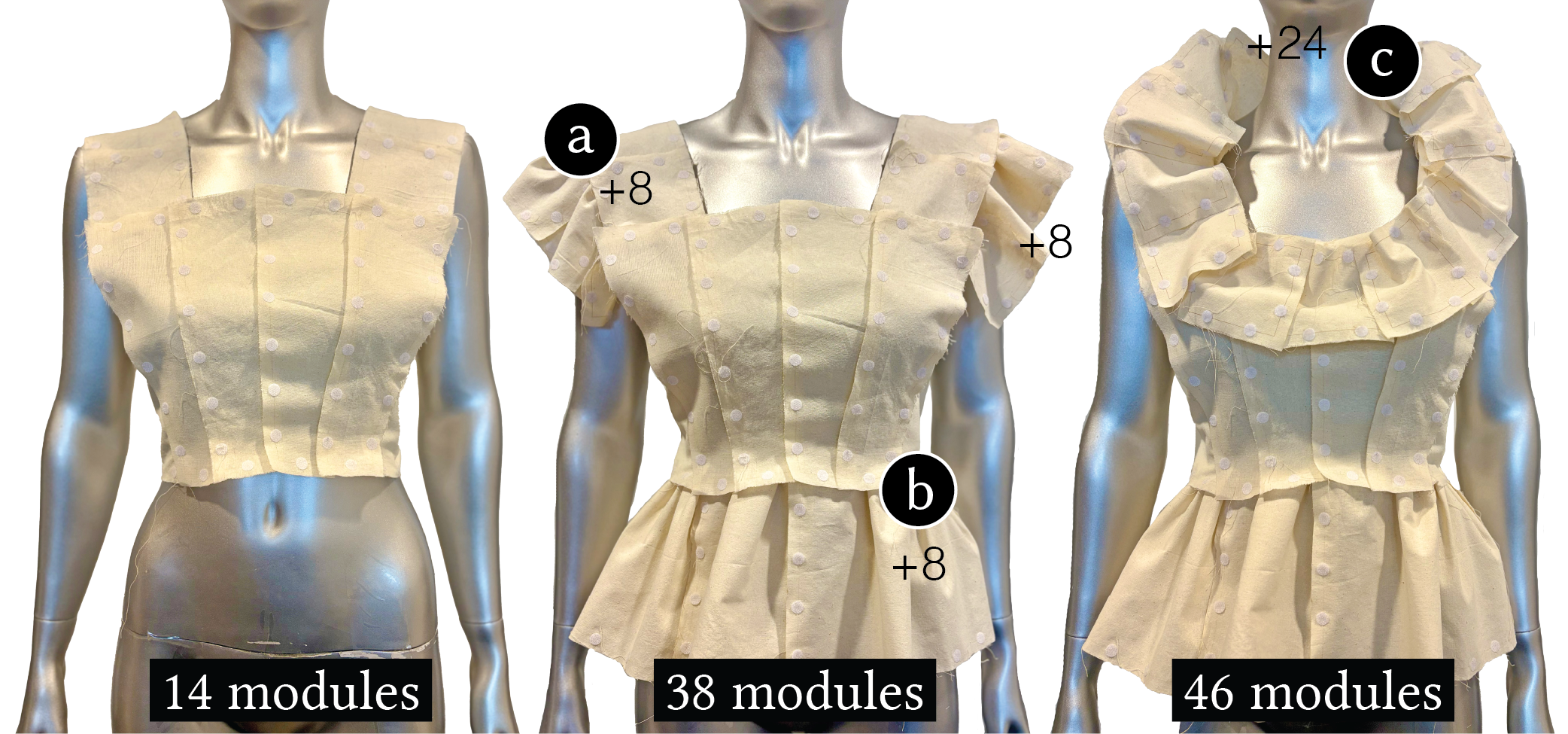}
\caption{\label{fig:add} Restyling a fitted, sleeveless top (a) in multiple ways: first by adding short ruffled sleeves (b) and a peplum (c), then reconfiguring the sleeves into a gathered collar.}
\Description{Progressive styling of a fitted top. Left: basic top. Middle: added ruffled sleeves and peplum. Right: swapped sleeves for gathered collar.}
\vspace{-0.2cm}
\end{figure}

Users can easily remix a base garment by adding components such as stylized sleeves, collars, or cuffs. \figref{fig:add} shows a fitted, sleeveless top first augmented with short ruffled sleeves and a peplum, then alternatively styled with a gathered collar—each variation adding distinctive flair to the base design. Users may also extend a garment. The trousers and dress in \figref{fig:teaser} are complete and wearable at multiple stages of construction (e.g., they can be worn as shorts or a top before additional modules are added).
This extensibility also enables detachability, useful in everyday contexts, such as shortening sleeves on a hot day.

\setlength{\columnsep}{10pt}
\begin{wrapfigure}{r}{0.26\linewidth}
  \vspace{-12pt}
  \begin{center}
\includegraphics[width=\linewidth]{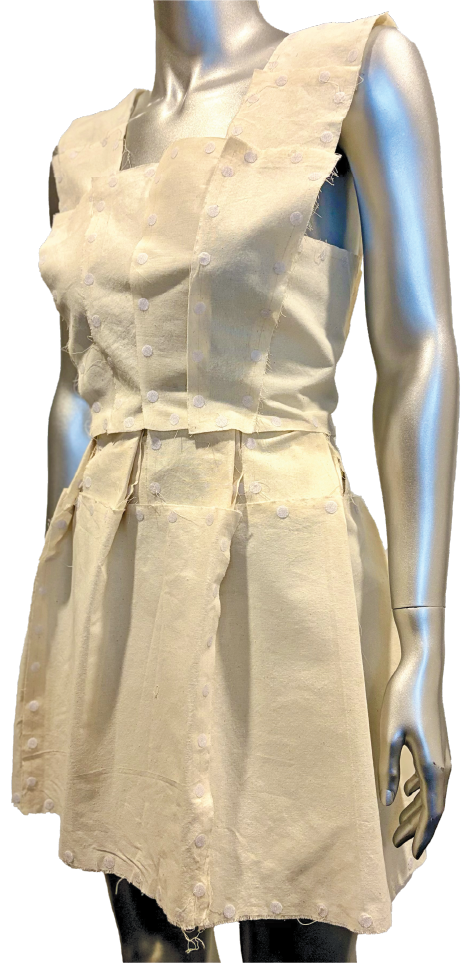}
  \end{center}
  \vspace{-10pt}
\end{wrapfigure}
\subsubsection{Combining Garments}\label{sec:combine}
Users can combine garments to create new designs. For example, the top from \figref{fig:add} can be merged with a pleated skirt to form the dress shown in the inset, or with pants for a jumpsuit.

\subsubsection{Transforming Between Garments}\label{sec:transform}
\begin{figure}[h]
\includegraphics[width=\columnwidth]{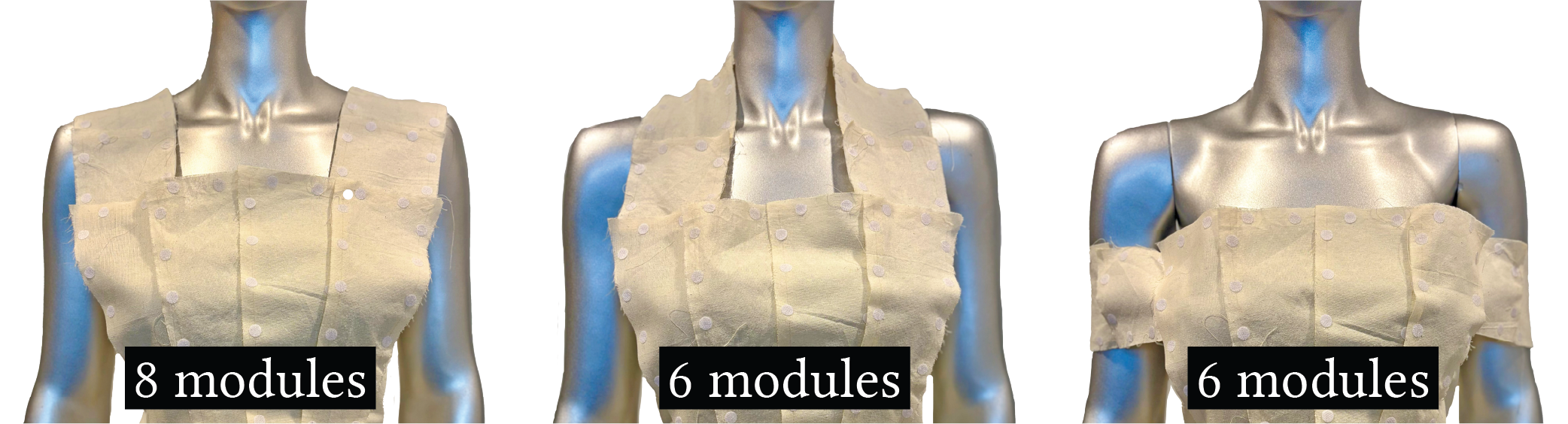}
\caption{\label{fig:local-reconfigure} Transforming a square neckline (left) into a halter-neck style (middle) and an off-the-shoulder look (right) by shortening the straps and adjusting their position. The module numbers specify only the ones needed for the straps.}
\Description{Strap reconfigurations. Square neckline, halter-neck, and off-shoulder styles achieved by altering strap positions.}
\end{figure}

Our system supports both local and global transformations. Local transformations involve rearranging components of a garment or modules within a single component (as shown in the straps \figref{fig:local-reconfigure}). Global transformations involve reconfiguring the entire garment, such as converting trousers into a dress (\figref{fig:teaser}).

\begin{figure}[h]
\includegraphics[width=\columnwidth]{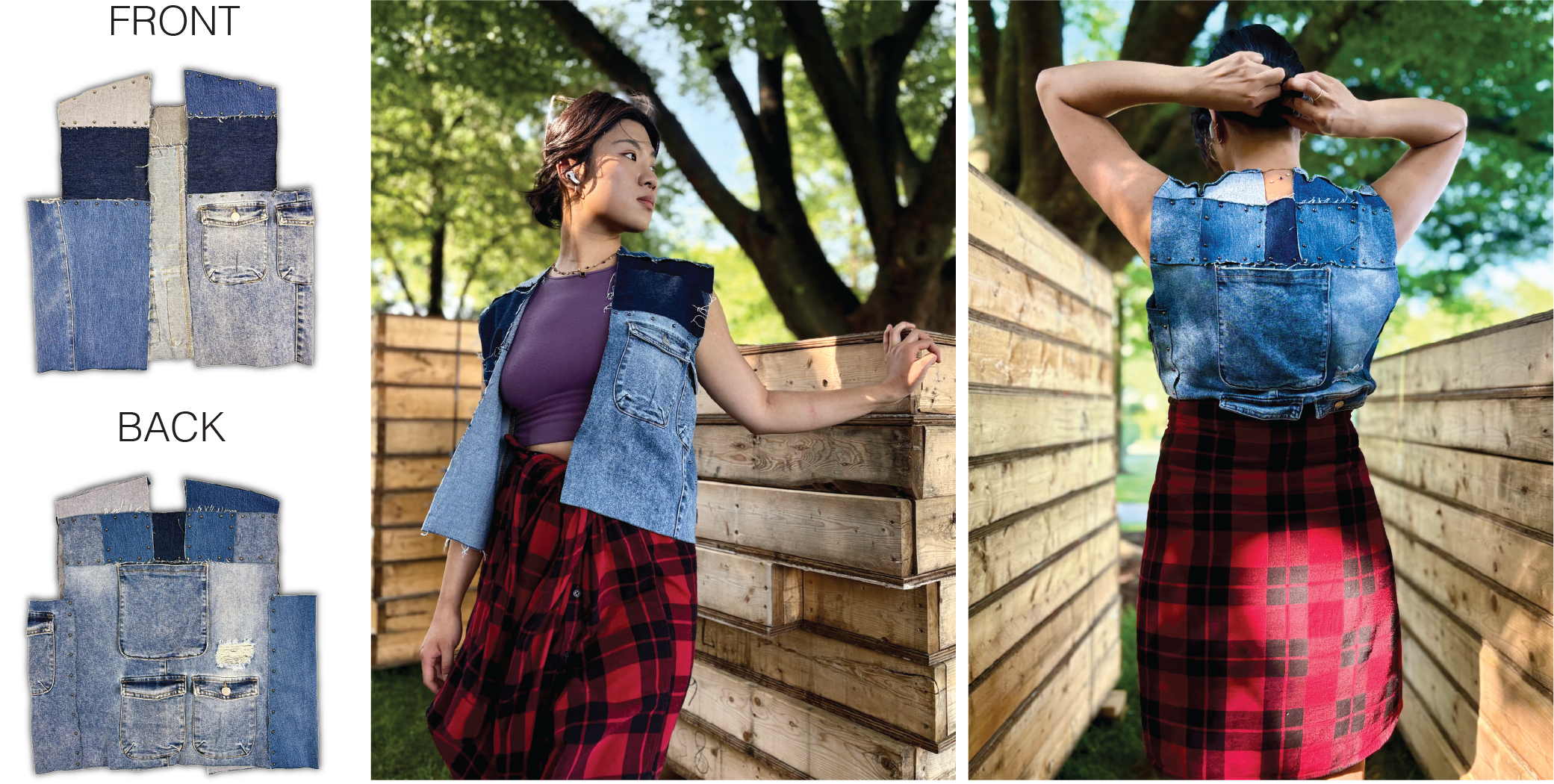}
\caption{\label{fig:repurpose}
\edited{A sleeveless denim jacket is made from modules cut from donated jeans, preserving original seams and pockets. The model wore it for 10 hours during a workday.}}
\Description{Outdoor wear test of denim jacket, cut and assembled from recycled jeans.}
\vspace{-0.3cm}
\end{figure}

\begin{figure}[h]
\includegraphics[width=\columnwidth]{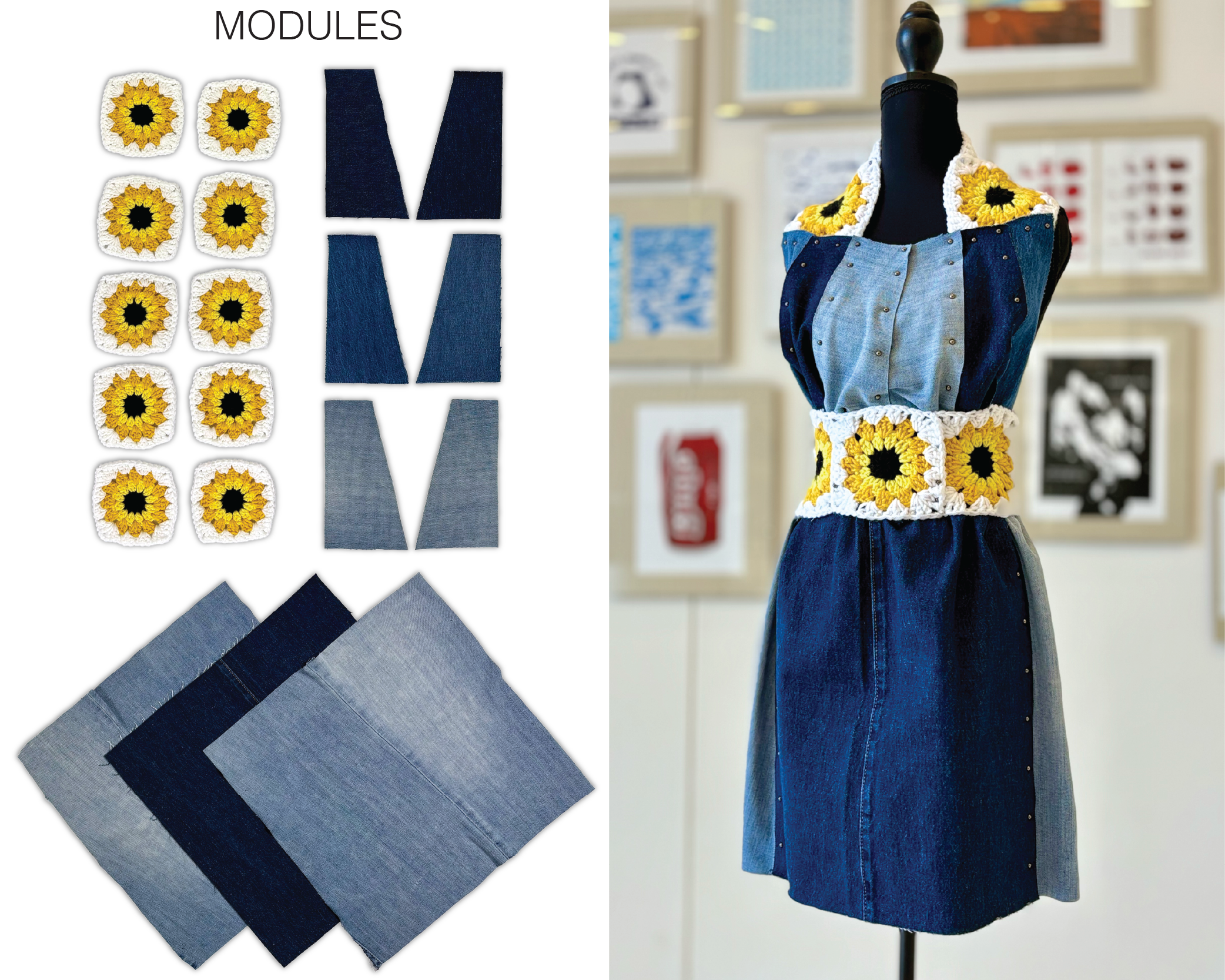}
\caption{\label{fig:patchwork} \edited{A backless halter-neck top is constructed from denim and crocheted granny square modules. A larger base unit size ($\Delta = 11$~cm) is used so the fabric modules match the dimensions of the existing granny squares.}}
\Description{Backless halter dress using mixed materials: denim squares and crochet granny squares; inset highlights module supply.}
\vspace{-0.2cm}
\end{figure}

\edited{\subsubsection{Textile Reuse}\label{sec:patchwork}
By supporting modules made from diverse materials, \garmod enables the creative reuse of existing textiles. The sleeveless jacket in \figref{fig:repurpose} is constructed from recycled jeans, while the backless halter dress in \figref{fig:patchwork} incorporates hand-crocheted granny squares to add distinctive styling to different parts of the garment.}

\begin{figure*}[b]
\includegraphics[width=\textwidth]{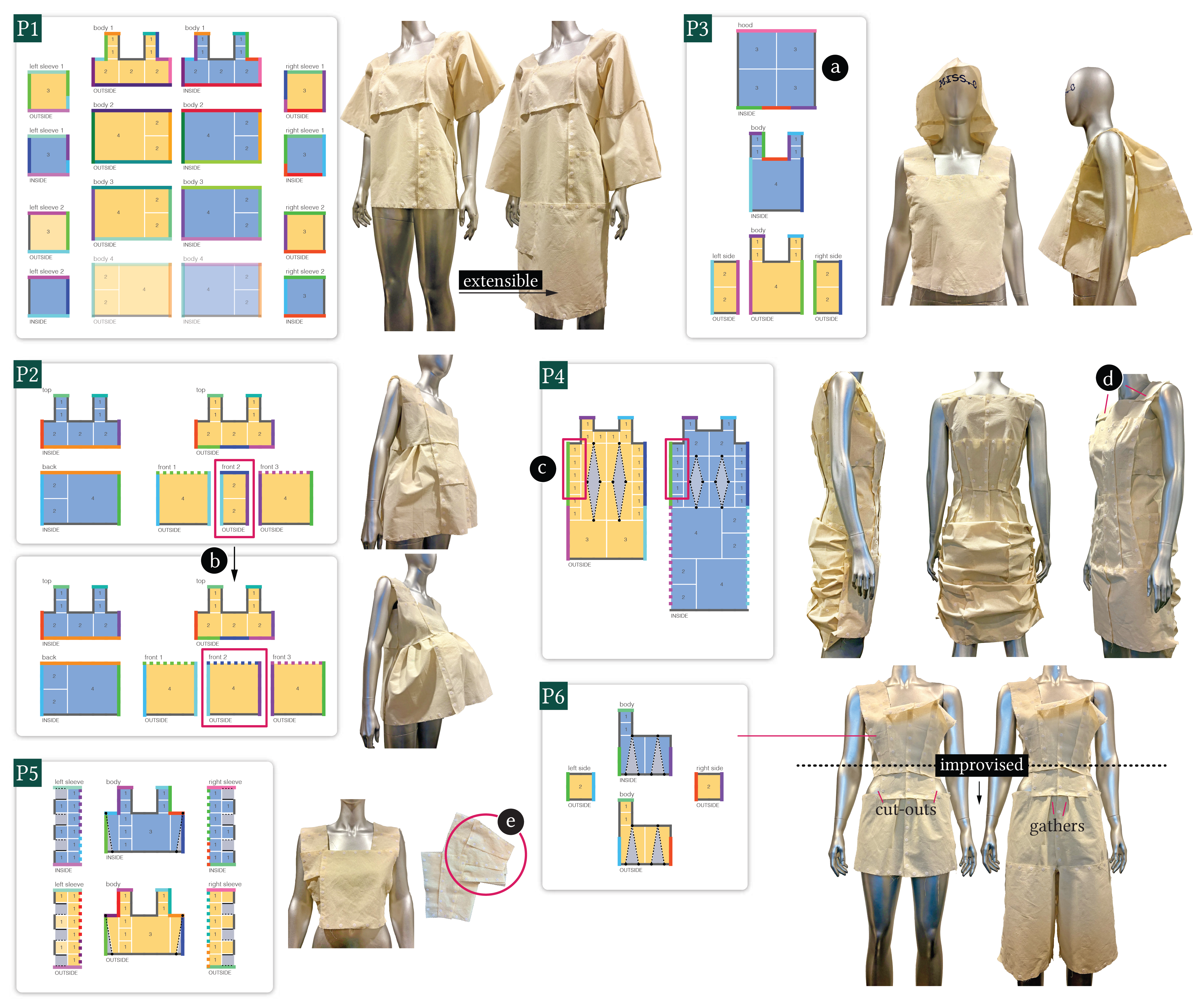}
\caption{\label{fig:user-designs}
Participants designed and assembled their own modular garment patterns.
P1 created a compound dress with extensible and detachable sleeves and skirt.
P2 designed maternity wear that accommodates two stages of pregnancy via a single panel swap (b).
P3 made a sleeveless hoodie inspired by the one they wore to the study session.
P4 designed a formal dress with horizontal gathers at the back of the skirt. We manually improved its assembly by replacing the $\D$-sized foundation modules along the sides (c) with larger $2\D$ units that span the seams. We also observed that increasing connection density along the seam interface (\secref{sec:seam-interface}) would yield cleaner joins than in the assembled prototype (d).
P5 sketched a top with gathered \textit{and} pleated sleeves, aiming to create a puff sleeve. However, due to a shortage of pre-fabricated modules, we created an ungathered version of the sleeve (e) instead.
P6 drew an asymmetric top and improvised the design into a jumpsuit during assembly. They added a waistband with cutouts and gathers to reconcile dimensional differences between the top and bottom.}
\Description{Showcase of six user-designed modular garments. Each includes a digital pattern with module layout and photos of the final garment. Designs range from dresses and hoodies to maternity wear and asymmetric jumpsuits.}
\end{figure*}

\subsection{User-Created Designs}\label{sec:userdesigns}
In our user evaluation, we invited participants to use \garmod~Studio to resize, restyle, and design their own modular garments.

\subsubsection{Participants}
We recruited six participants (mean age 28) from makerspaces at a local university. All had prior sewing experience, including making alterations and sewing garments from existing patterns. Three participants (P2, P4, and P6) had experience designing their own patterns. All had worked with CAD software (e.g., SolidWorks), and five (all except P5) were proficient in SVG drawing tools (e.g., Illustrator). \edited{P2 and P6 had decades of design experience from architectural training and managing makerspaces.}

\vspace{-0.1cm}

\subsubsection{Study Procedure}
Each session lasted approximately two hours and involved three main tasks using \garmod~Studio:
\begin{enumerate}
\item \textbf{Resize a T-shirt (15 min):} Participants resized a given T-shirt pattern to fit a different body model as in \figref{fig:resize}.
\item \textbf{Add a Shaping Feature (15 min):} Participants selected a pattern and incorporated at least one garment feature.
\item \textbf{Design a Garment (15 min digital design + 30 min physical assembly):} Participants designed a garment from scratch and constructed their designs from pre-fabricated prototyping modules guided by system-generated assemblies.
\end{enumerate}
One participant (P2) returned later to create additional garments.

\subsubsection{Task 1: Resizing a T-shirt}
Participants redrew a T-shirt pattern to accommodate new target measurements (\figref{fig:resize}). Resizing was considered successful if the modified pattern fit correctly on the specified body model in simulation. All participants completed the task successfully within 15 minutes. While some found drawing and stitching the armhole edges somewhat tricky, all participants rated the overall task as ``very easy'' or ``easy'' on a 5-point Likert scale. For P2, \participantquote{solving the puzzle} was their favorite part.

\subsubsection{Task 2: Adding a Garment Feature}
Participants selected and augmented a base garment by adding at least one feature—gathers, pleats, or darts. Some (P4, P5) recreated the compound skirt from \figref{fig:walkthrough}, while others (P2, P3) explored more open-ended manipulations, such as gathering a single sleeve or repeatedly adding pleats at seams (e.g., \figref{fig:operation-rematch}). Four participants (P1, P3, P4, and P5) rated this task as ``easy,'' while P2 and P6 found it ``slightly difficult,'' primarily due to the challenge of understanding more complex feature interactions (\appref{app:feature-casework}) and how they translate to fabric manipulation in the physical world.

\subsubsection{Task 3: Designing a Garment}
Participants designed and assembled a modular garment from scratch.

\textbf{P1} created a compound dress with extensible sleeves and body (\figref{fig:user-designs}a), designed to adapt to changing temperatures throughout the day (e.g., transitioning from short to long sleeves in the evening).

\textbf{P2} designed a piece of maternity wear featuring an interchangeable panel to accommodate different stages of pregnancy (\figref{fig:user-designs}c). They used gathers in the midsection to provide additional space while keeping the top of the garment unchanged. The garment was constructed in under 15 minutes and later modified in less than 5 minutes, demonstrating the efficiency of modular construction for both initial assembly and subsequent reconfiguration to achieve significant fit adjustments.

\textbf{P3}, after experimenting with simulation features, designed a sleeveless hoodie inspired by the one they wore to the study session (\figref{fig:user-designs}c). Although reverse-engineering the hood’s pattern took some time, recreating and validating it in the studio was quick (15 min), as was the physical construction (10 min).

\textbf{P4} designed a dress with a skirt gathered horizontally at the back (\figref{fig:user-designs}d). They iterated multiple times on the placement of the gathers, using the simulation to refine their design. P4 found the unit-based system helpful for quick measurements and counting (e.g., the widest part of the mannequin had a circumference of 12$\D$), which simplified the design process compared to traditional sewing methods. Due to the complexity of the gathers, assembly took approximately 30 minutes.

\textbf{P5} created a top featuring gathered and pleated sleeves (\figref{fig:user-designs}e), aiming for a short puff-sleeve design. To achieve the desired fullness and shaping, they combined gathers and pleats in parallel.

\textbf{P6} began with an asymmetric top, aiming to challenge the more regular nature of our modular garments, and later improvised complementary pants to form a jumpsuit (\figref{fig:user-designs}f). The improvisation demonstrated both the flexibility of our modular system and the value of digital prototyping. To resolve the size mismatch when joining the top and bottom, the participant added a waistband of unit-sized modules with cutouts and gathered the pants to attach them. The top, based on a pre-designed digital pattern, took just 10 minutes to assemble, while the improvised pants required 40 minutes of trial and error. However, the arrangement of small modules and cutouts along the waistband inspired the participant to suggest a patchwork version if they created a higher-quality version of the garment from the muslin prototype.

\subsection{User Feedback}
We invited participants to provide feedback on their experience designing and assembling modular garments with \garmod.

\subsubsection{Adaptation}
Participants highlighted the value of modular garments in adapting to evolving environments, bodies, and needs. P5 noted the garments' flexibility in responding to changing weather conditions and facilitating the shift from professional to casual attire. P2 described modularity as ``ideal'' for maternity wear, appreciating the ability to adjust rather than replace an entire wardrobe. They also pointed to children's apparel as a promising application, given how rapidly kids outgrow clothes. Looking beyond garments, they envisioned accessories like bags that could \participantquote{change size depending on how much stuff you need to carry that day}.

\subsubsection{Rapid Prototyping}\label{sec:prototyping}
Participants valued \garmod's support for rapid prototyping. P1, P3, and P4 appreciated being able to iterate quickly digitally, guided by feedback from the simulation. P3 emphasized the efficiency of assembling mockups from simple, reusable modules, expressing enthusiasm for fast, low-cost iteration without the worry of fabric waste. P4 highlighted the ease of thinking in terms of modular ``units'' for both sizing and styling, as well as the ability to validate a concept digitally and physically without the need for cutting or sewing. P1 echoed this sentiment, noting that they would use the system for general pattern-making by rapidly prototyping a garment's baseline silhouette and style before fine-tuning the pattern.

\subsubsection{Sewing-Free Construction}
Participants emphasized that by abstracting and simplifying traditional garment-making, \garmod~is more approachable for those with limited sewing experience.
P1 noted that the system could serve as an effective learning tool for garment construction—even beyond modular designs—by simplifying elements like seams, gathers, and pleats.
Several participants appreciated that \garmod~enabled beginners to quickly produce functional garments; for example, P4 said, \participantquote{I learned [the system], made the pattern, and assembled it in two hours.}
P4 further explained that, unlike traditional methods involving extensive sketching, proportion calculations, and multiple test pieces, \garmod's streamlined approach simplifies the entire process.
P3 likened its modeling experience to \participantquote{a Tinkercad aesthetic,} suggesting it could serve a similar role as an approachable and playful platform for rapid garment design and learning.

\subsubsection{Modularity as an Aesthetic}
Participants expressed enthusiasm for modularity not only as a functional feature but also as an aesthetic choice. P3 posed the question of whether modular clothing should conceal its structure or instead highlight it to \participantquote{purposefully make a statement.} P2 advocated for embracing a bold modular aesthetic, envisioning garments that \participantquote{push the box aesthetic} as a counterpoint to fleeting fashion trends. P6 was particularly excited about using the system’s constraints to explore visual patterns such as patchwork and cutouts. P2 also expressed interest in adding functional or decorative accessories—such as pockets and bows—either to existing connection points or via additional hardware.

\section{Discussion} \label{sec:discussion}

Many insights from our user evaluation highlight broader themes related to designing for reconfiguration and reuse within fashion and across other domains.

\subsection{Designing for Re-Assembly}
To reduce assembly time, our method minimizes the number of modules required to construct a design, often favoring larger modules when possible. However, larger modules can restrict fabric variation within a region—limiting visual expression when modules are used as design features (e.g., \figref{fig:patchwork})—and may increase the effort required to reconfigure garments, as fewer, larger blocks offer less flexibility for reshaping or rearranging. Several participants (P1, P2, P5) observed that fully disassembling and reassembling a garment is often unnecessary when transitioning between modular designs. For instance, clusters of modules forming larger panels—such as those forming a bodice or skirt—can often be reused with little to no breakdown. This insight suggests opportunities for ``capsule panels'' or guided assemblies that preserve and reuse multi-module groups. It may also be beneficial to design multiple modular garments as a cohesive set, selecting module sizes and materials to support efficient and expressive recombination.

\subsection{Designing for Change}
Our modular garments are intended to be reconfigured over time. While a wearer might switch between predefined styles—such as adding or removing a hood, or converting pants into shorts—it remains unclear how often such reversible transformations occur in practice. Alternatively, one could imagine creating a new configuration each day, never repeating the same garment form. These varied use cases suggest that design strategies should consider not only the end configurations but also the frequency and nature of change. Depending on how often and in what ways a garment is expected to be reconfigured, transitions can be optimized—for example, by designing specific transformation pathways and minimizing the steps or modules needed to move between them.

\subsection{Simple Modules Promote Efficiency}

The regular geometries of the modules significantly improve material efficiency compared to conventional garment patterns, whose irregular contours often leave large amounts of unused fabric when cut from fabric sheets, and their uniform and repetitive design streamlines assembly. Participants were able to construct garments by following consistent attachment patterns without the need for individualized instructions for each piece.
Our set of modules, though small, achieved a wide range of shaping by supporting gathers, pleats, and darts through carefully designed interfaces and specialized blocks. The shaping modules, though more complex than the foundational components, efficiently transform 2D sheets into structured 3D forms that fit the body and remain reusable across a variety of garment sizes and styles.
Extending this set of modules to allow for even more customizable sizing and closures, such as drawstrings or zippers, would further bring modular clothing closer to the fits achieved by conventional clothing.

\section{Limitations \& Future Work}
Our definition of modular garments represents just one possible interpretation of modularity. While our formulation of the design space supports a range of garments and use cases, there are many opportunities to extend these ideas and improve the system.

\subsection{Sizing Control}
While users can customize module dimensions, the unit size $\D$ controls the granularity of fit adjustments (e.g., with $\D=8~cm$, all sizing occurs in 8~cm increments). We observe a trade-off between unit size and the number of modules—and thus the likely assembly time—required for a given garment.
Although participants achieved a wide range of garment fits using our modules, more precise fitting could be supported by offering a more continuous range of module sizes or in-module adjustments. For example, flexible interfaces, such as drawstrings or laces, could enable fine-grained control. Finally, alternative module shapes (e.g., curved pieces) may help modular garments better emulate conventional clothing.

\subsection{Supporting Multi-Fabric Designs}
Several participants (P2, P3, P6) discussed embracing modularity by making the units more visually distinct through patchwork. Using modules in different colors, prints, and textures could give wearers greater control over a garment’s appearance while highlighting its modular structure. Future work could explore interactive decomposition tools that allow users to place modules dynamically via a drag-and-drop interface to achieve custom patterns.

\subsection{Wearability}\label{sec:wearability}
While participants enjoyed designing and assembling modular garments, some (P2 and P3) raised concerns about longer-term wearability. P2 joked, \participantquote{If I bent over and the snaps unsnapped, that would be super embarrassing}. \edited{In response, the first author wore the sleeveless denim jacket from \figref{fig:repurpose} over the course of a day across various settings, including the office, machine shop, and a dinner outing with friends without noticeable decline in quality.
Nevertheless, further work is required to ensure these garments meet the standards of comfort and durability required for daily wear and regular washing~\ref{goal:comfortable}.}
\section{Conclusion}\label{sec:conclusion}
Modular garments offer a pathway toward clothing that adapts to evolving styles, fits, and needs—while supporting repair, reuse, and self-expression.
\garmod~demonstrates one approach to designing such dynamic garments.
We hope this work inspires future efforts toward a more sustainable and creative future for fashion.

\begin{acks}
We thank Kevin Zhang and Yunhan Dai for assistance with fabrication, Benjamin Chen and Esther Lin for feedback, and Erik D. Demaine for mentorship. This work was supported in part by the MIT Morningside Academy for Design (MAD), an MIT MAKE Design-2-Making Mini-Grant, and an NSERC PGS-D scholarship. We gratefully acknowledge the use of facilities at the CSAIL Machine Shop, the CBA Fab Lab, and the Future Sketches group at the Media Lab.
\end{acks}

\bibliographystyle{ACM-Reference-Format}
\bibliography{modularclothing}

\clearpage

\appendix
\section{Feature Casework}\label{app:feature-casework}

We provide detailed casework outlining the types of user interactions and control available when manipulating gathers, pleats, and darts in \garmod Studio.

\subsection{Gathers}\label{app:gather_details}
\begin{figure}[h]
\includegraphics[width=\columnwidth]{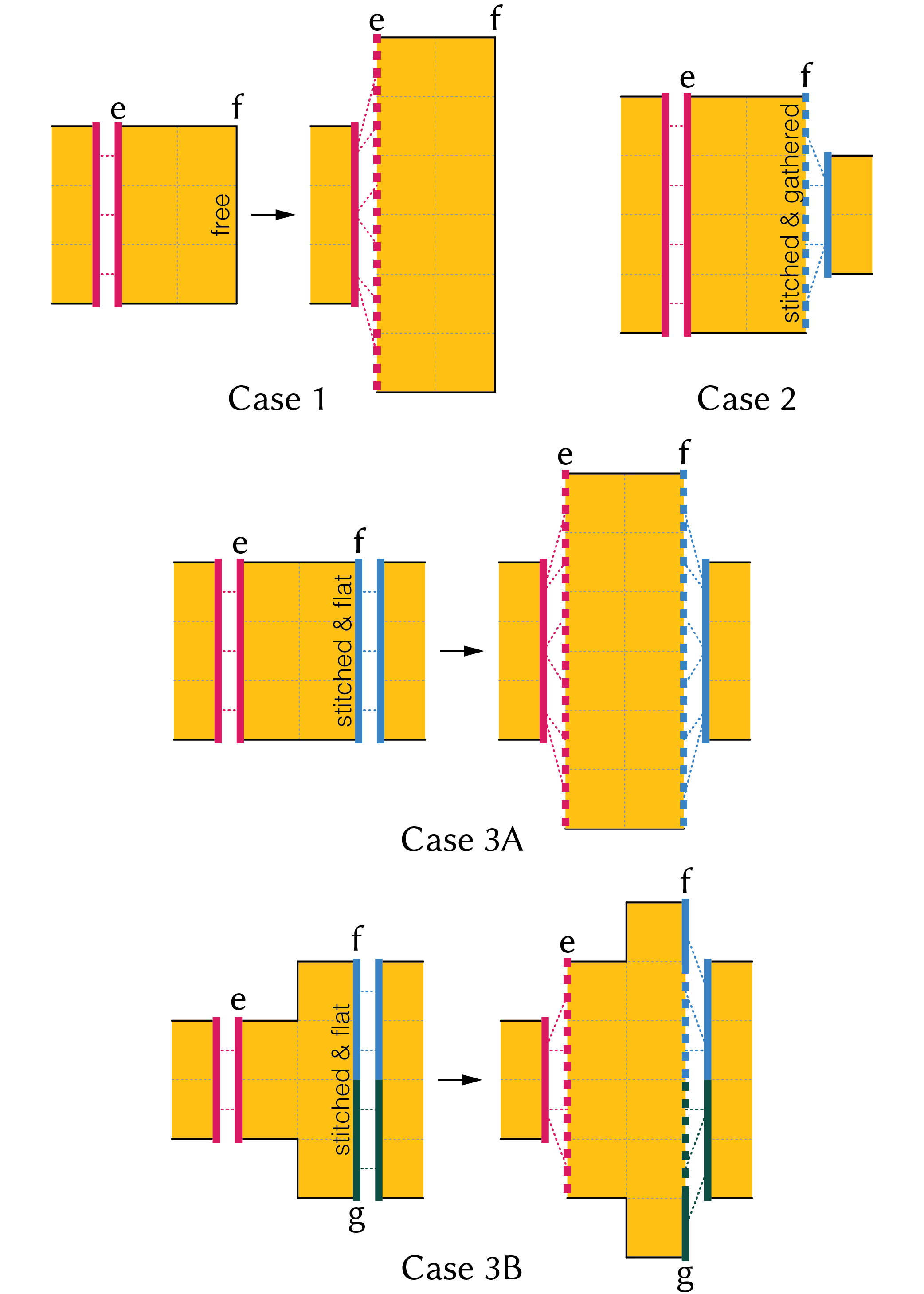}
\caption{\label{fig:gather-cases}
Scenarios for gathering edge $e$ in the red seam,
depending on the state of its opposite edge $f$.
Case 1: If $f$ is a free edge, both $e$ and $f$ double in length;
Case 2: If $f$ is already gathered, then gathering $e$ would exceed $f$'s maximum length and is therefore prohibited;
Case 3AB: If $f$ is part of a flat seam, both $e$ the segments of $f$ \emph{directly opposing} $e$ are gathered.
Dotted lines indicate edge correspondences at the unit-segment-level.
}
\Description{Seam gathering scenarios. Diagrams show how gathering one edge affects the opposing edge.}
\end{figure}

The user initiates a gather by selecting an edge along a seam.
Gathering doubles the length of the selected edge, and to maintain orthogonal panel geometry, the opposing edge must also double in length.
\figref{fig:gather-cases} illustrates several scenarios that show whether and how the panel and surrounding seams are adjusted to accommodate the gathering.

By default, our system minimizes cascading changes across panels.
In Case 3A, for example, we gather $f$ rather than expanding the adjacent panel it connects to.
If the user prefers to keep $f$ flat and instead double the connected panel, we provide interactions to support this alternative adjustment (\figref{fig:convert-to-pleat}c).

\subsection{Pleats}\label{sec:pleat_details}
\begin{figure}[h]
\includegraphics[width=\columnwidth]{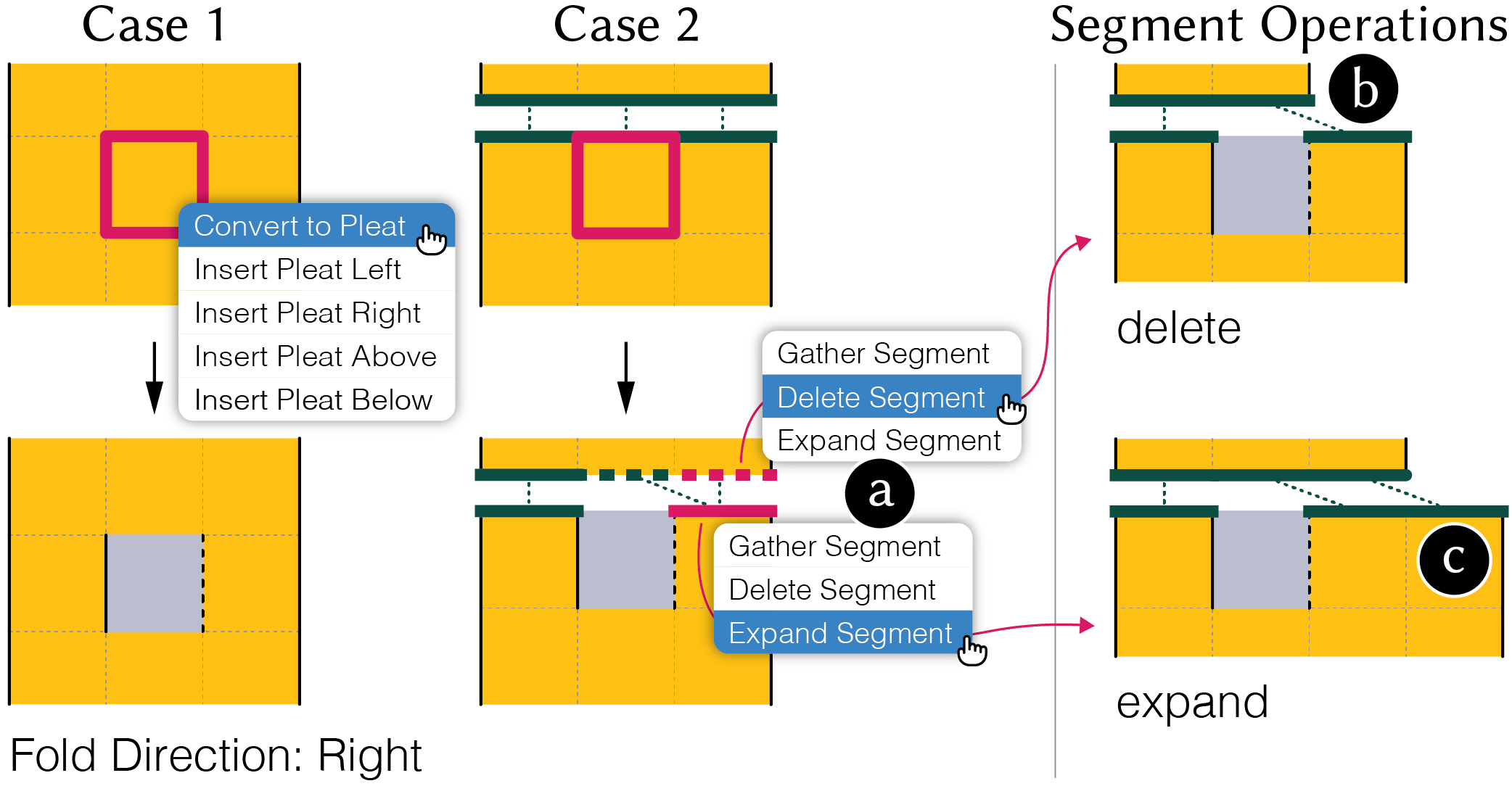}
\caption{\label{fig:convert-to-pleat}
Converting a panel unit into a pleat results in one of two scenarios.
Case 1: If the pleat is not adjacent to a seam \emph{or} its fold direction is not orthogonal to the seam edge, no seam adjustments are needed.
Case 2: If the pleat is flush against a seam \emph{and} folds orthogonally—removing a unit edge segment—the opposing segment must be rematched, introducing a gather (a).
Users can resolve undesired gathers by either
(b) deleting the gathered segment on the opposite edge or
(c) expanding a unit segment on the current edge to preserve the original seam length after folding.}
\Description{Pleat conversions. Case 1: internal cell conversion leaves seam unchanged. Case 2: boundary cell conversion introduces seam gathers with two resolution options.}
\end{figure}

\begin{figure}[h]
\includegraphics[width=\columnwidth]{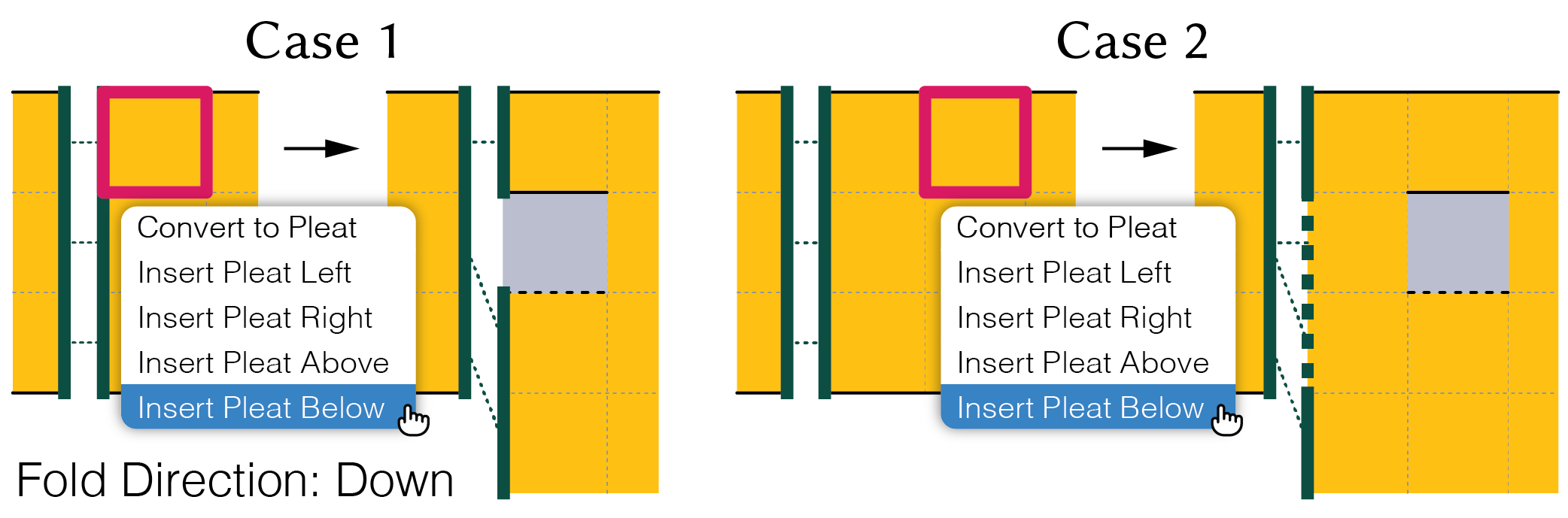}
\caption{\label{fig:insert-pleat}
Inserting a pleat introduces a new unit-width strip of geometry to the panel. This may affect seams at the strip's endpoints—as seen in Case 2, where gathering is required.
However, if the pleat is adjacent and folds perpendicular to the seam—removing the newly introduced edge length—then no rematching is needed, as shown in Case 1.}
\Description{Pleat insertion effects on panel geometry and seam matching. Case 1: insertion along an edge keeps seam matches unchanged. Case 2: internal insertion alters seam matches.}
\end{figure}

To pleat, the user first specifies a fold direction—right, left, up, or down—and then either converts a selected panel unit into a pleat module or inserts a pleat module adjacent to it.
The former may remove a unit-length edge through folding, requiring seam adjustments (e.g., Case 2 in \figref{fig:convert-to-pleat}).
The latter can preserve edge dimensions after folding (e.g., Case 1 in \figref{fig:insert-pleat}) but alters the geometry of the panel.
We illustrate the possible interactions for each operation in the following two figures, respectively.

\subsection{Darts}\label{sec:dart_details}
To add a dart, the user clicks any grid point within a panel—defining the midpoint of the cut-out’s triangular base—and selects an orientation (vertical or horizontal) from the dropdown that appears. Dart height and width are chosen from a separate dropdown above the editor. A dart is added only if there is sufficient space in the panel and the operation respects constraints along the seams.
These considerations are illustrated in the following two figures.

\begin{figure}[h]
\includegraphics[width=\columnwidth]{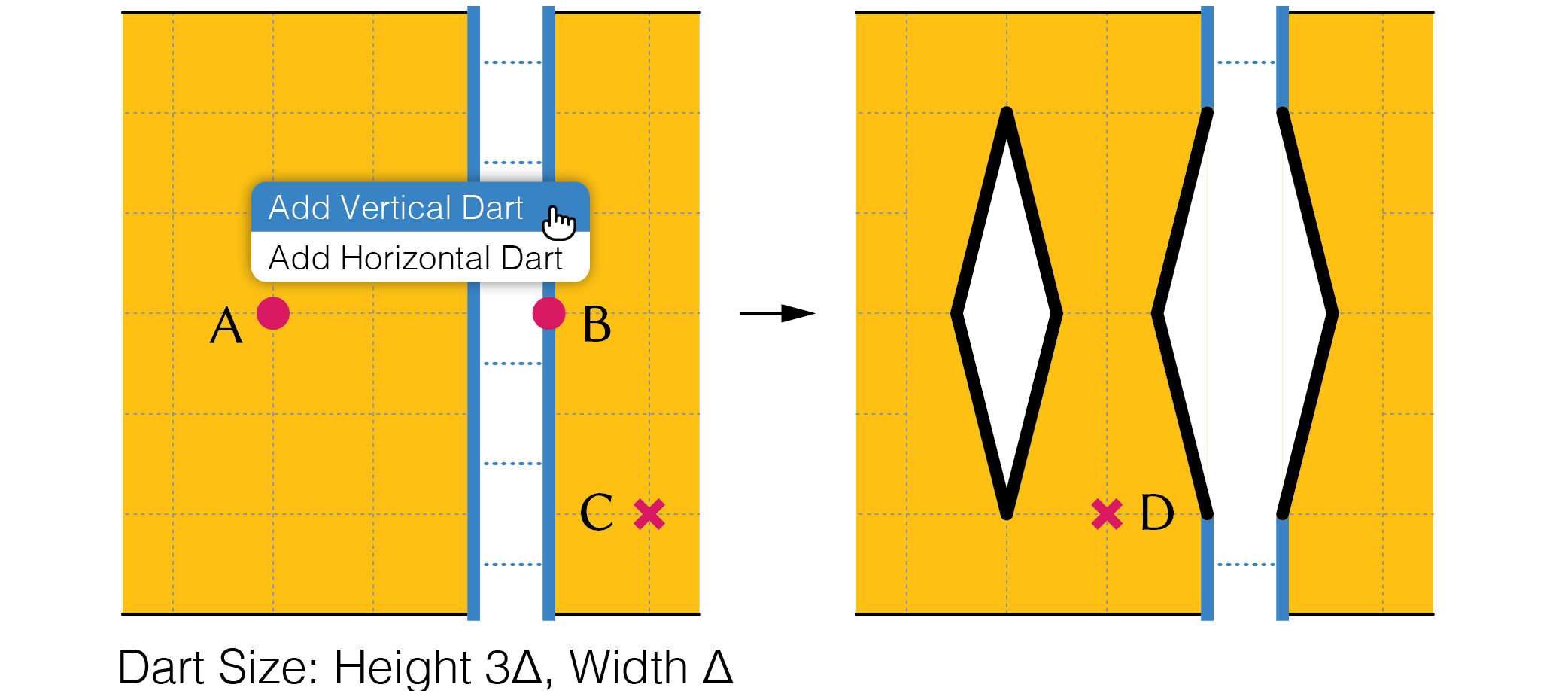}
\caption{\label{fig:diamond-dart-case}
When a dart is added at an interior point of a panel, it is always a diamond dart—formed by aligning two dart modules—and the system checks only for sufficient space.
Consider adding vertical diamond darts at points A, B, C, and D in sequence:
A succeeds, as the dart fits entirely within the panel.
B lies along a seam; since there is space across both panels, the dart is allowed.
C is disallowed, even without B, as it would extend beyond the panel boundary.
D is also disallowed, as it would intersect with the dart at B.}
\Description{Examples of valid and invalid dart placements. (A) Interior diamond dart. (B) Edge-aligned dart pair. (C) Over-boundary dart (invalid). (D) Intersecting dart (invalid).}
\vspace{-0.5cm}
\end{figure}

\begin{figure}[h]
\includegraphics[width=\columnwidth]{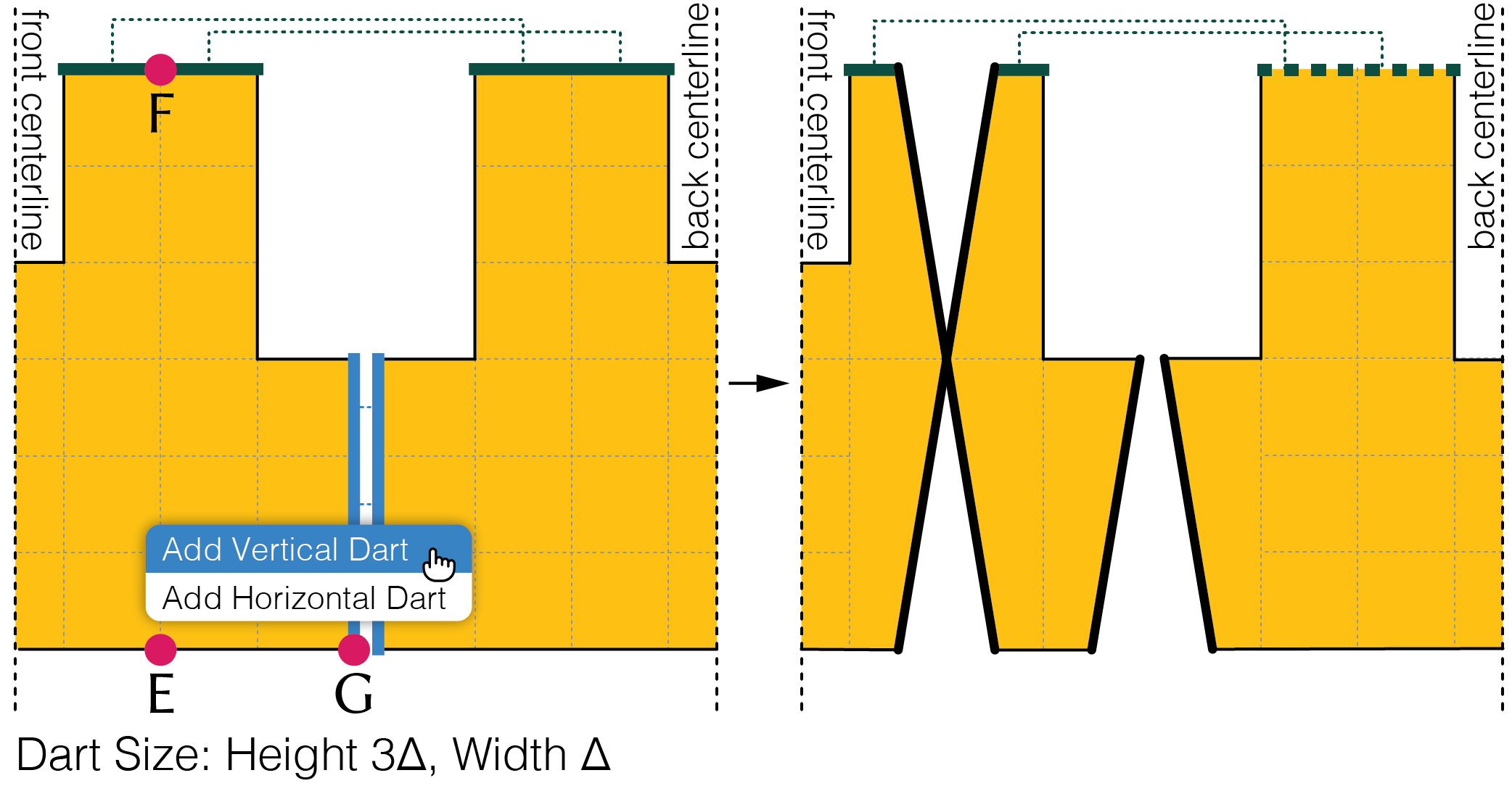}
\caption{\label{fig:triangle-dart-case}
Adding darts along the edges of a panel involves more complex casework, often requiring coordination with seam constraints.
E: A dart is added perpendicular to a free edge;
F: A \emph{universal} dart (with width \D) is added perpendicular to a seam that is not already gathered;
G: A dart is added at a corner, with its two parts joining across a seam.}
\Description{Edge-specific dart applications. (E) Perpendicular dart on a free edge. (F) Universal dart on an ungathered seam. (G) Corner dart split across seams.}
\end{figure}

\section{3D Simulation}\label{app:simulation}

We implement the 3D garment simulation in \garmod~Studio using Blender’s open-source mass-spring cloth model.

\subsection{Feature Modeling}
To simulate fabric manipulations such as gathers and pleats, we modeled them directly as follows.

\begin{figure}[H]
\includegraphics[width=\columnwidth]{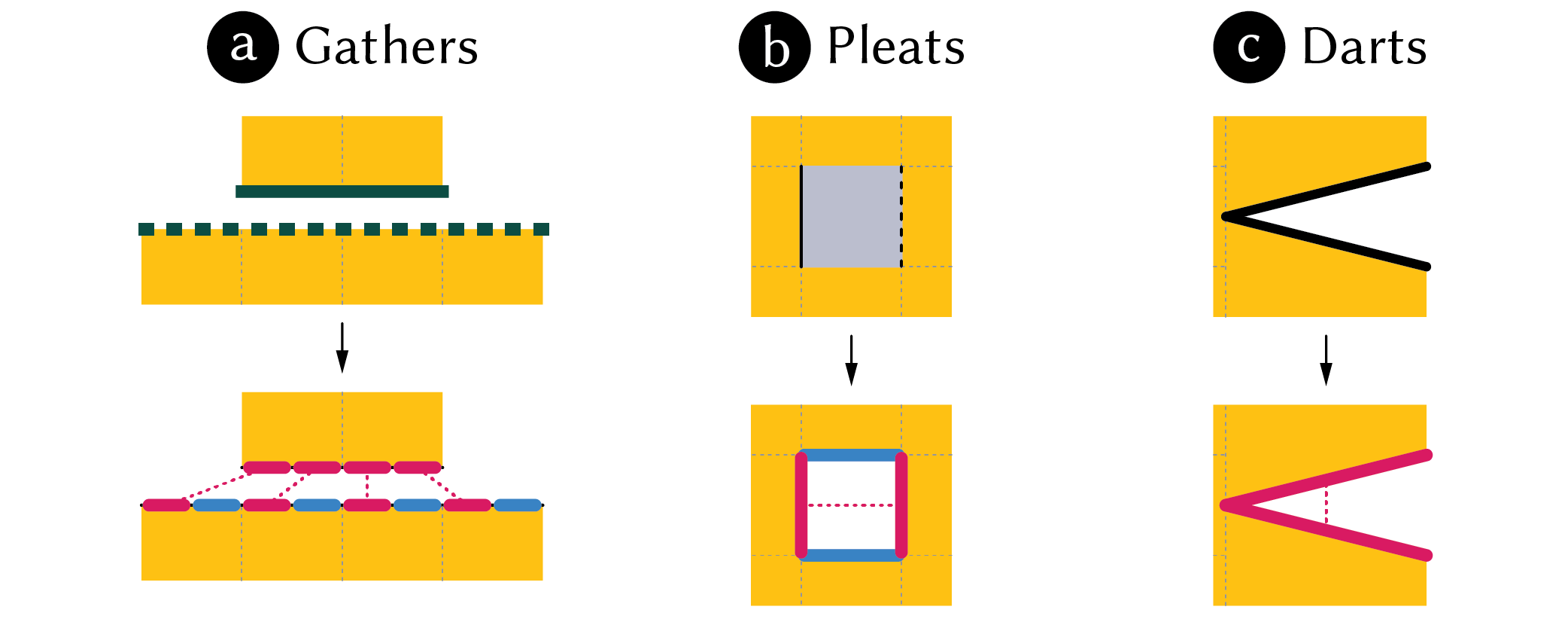}
\caption{\label{fig:feature_modeling}
Modeling gathers, pleats, and darts for simulation.
Red segments connected by dotted lines are matched and sewn together during simulation. Blue segments are self-matched and fold in half during simulation.}
\Description{Simulation representation of shaping features: gathers, pleats, and darts.}
\end{figure}

For gathers, we match seams at the \emph{half}-unit level, as shown in \figref{fig:feature_modeling}a. Each half-unit segment along the flat edge is matched (and later stitched to) every other half-unit segment on the longer edge (red), while the remaining unmatched segments on the longer edge self-match to fold in half during simulation.

For pleats, we treat the modules as cut-outs or holes in the fabric as shown in \figref{fig:feature_modeling}b. Edges that align during folding are explicitly matched (red), while remaining edges self-match (blue). For pleats along panel edges, their matches may span adjacent panels. Darts are similarly treated as cut-outs, with their legs matched and sewn together during simulation (\figref{fig:feature_modeling}c).

\subsection{Meshing}

\setlength{\columnsep}{10pt}
\begin{wrapfigure}{r}{0.42\linewidth}
  \vspace{-12pt}
  \begin{center}
    \includegraphics[width=\linewidth]{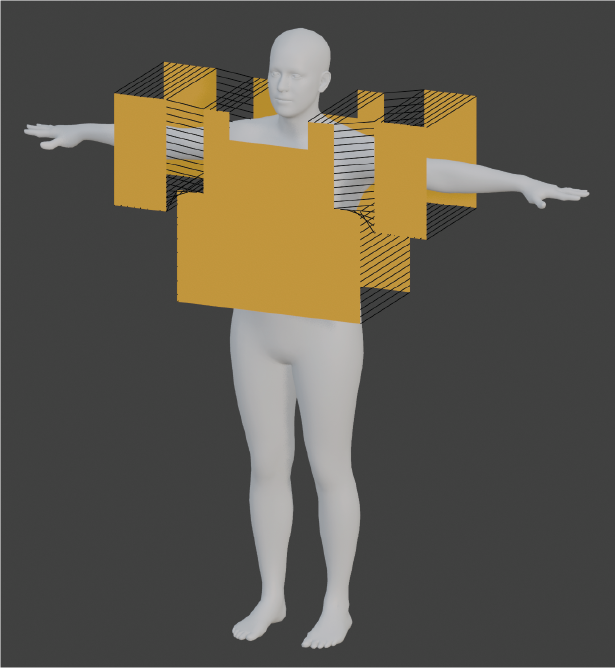}
  \end{center}
  \vspace{-14pt}
\end{wrapfigure}
Given a pattern represented as described above, we prepare a mesh as follows.
First, we convert coordinates from the pattern alignment view (\figref{fig:system_overview}) to Blender's coordinate system. Panels marked as ``outside-up'' are placed in front of the body model (+0.2~m), and ``inside-up'' panels behind it (–0.2~m).
Next, we subdivide each panel edge by inserting Steiner points at regular 1~cm intervals, then triangulate each panel using the Triangle library~\cite{shewchuk1996triangle}.
Finally, we add ``sewing threads''—spring-based edges—between corresponding vertices of matching panel edges (see inset).
Edge subdivision controls both triangle size and the density of sewing threads, which pull the panels together during simulation to wrap the garment around the body model.

\subsection{Results}

We approximate a medium-weight woven fabric using Blender’s cloth modifier, with balanced stiffness—tension, compression, shear, and bending all set to 3—and an areal density of \(0.6~\mathrm{kg/m^2}\), which considers the additional weight contributed by garment connectors.

We use Blender’s default cloth solver with a moderate simulation quality setting (7 substeps per frame). The simulation is run for 100 frames and typically converges within 30, requiring only 3–5 seconds on our MacBook Air, M2, 2023, with 8~GB memory.

\begin{figure}[h]
\includegraphics[width=\columnwidth]{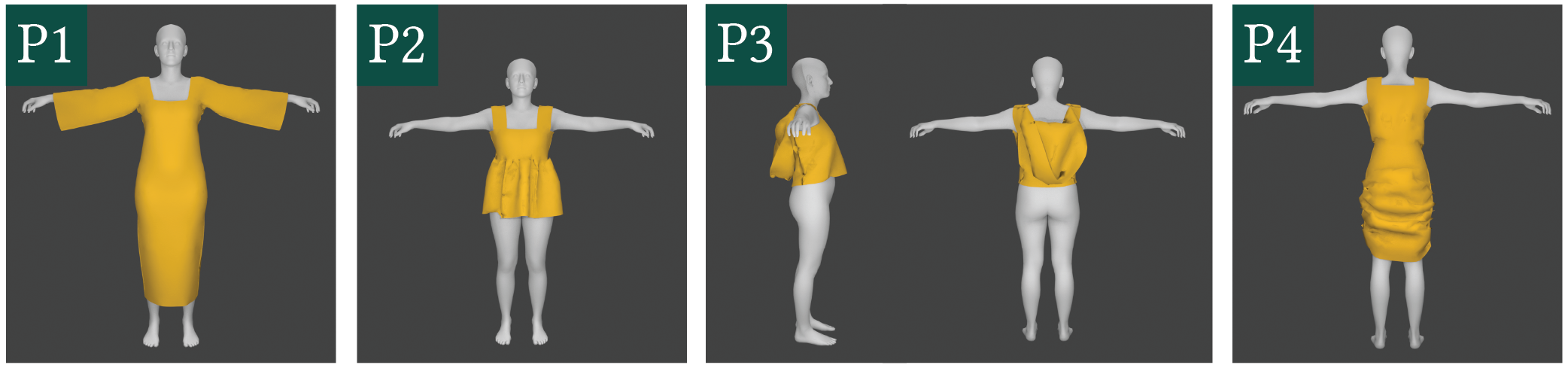}
\caption{\label{fig:simulation_results}
Screenshots of simulation results from our user study, showcasing a range of garments with varying features and complexities on different avatars.}
\Description{Garment simulation results. Avatars of two different sizes wearing user-designed garments with varied shaping features.}
\end{figure}

\section{Decomposition Evaluation}\label{apx:decomp_eval}

We conducted a brief evaluation of our ILP-based decomposition approach to understand how pattern characteristics—specifically area, irregularity, and panel count—affect solver runtime and assembly efficiency, measured by the number of modules. All experiments assumed unlimited availability of module sizes from $1$ to $4$ (i.e., $\Delta=1$). We used Google Colab (Python~3 on a Google Compute Engine backend, with 12.7~GB of RAM). While preliminary, the results demonstrate sufficient performance for many practical use cases. 

\begin{figure}[h]
\includegraphics[width=\columnwidth]{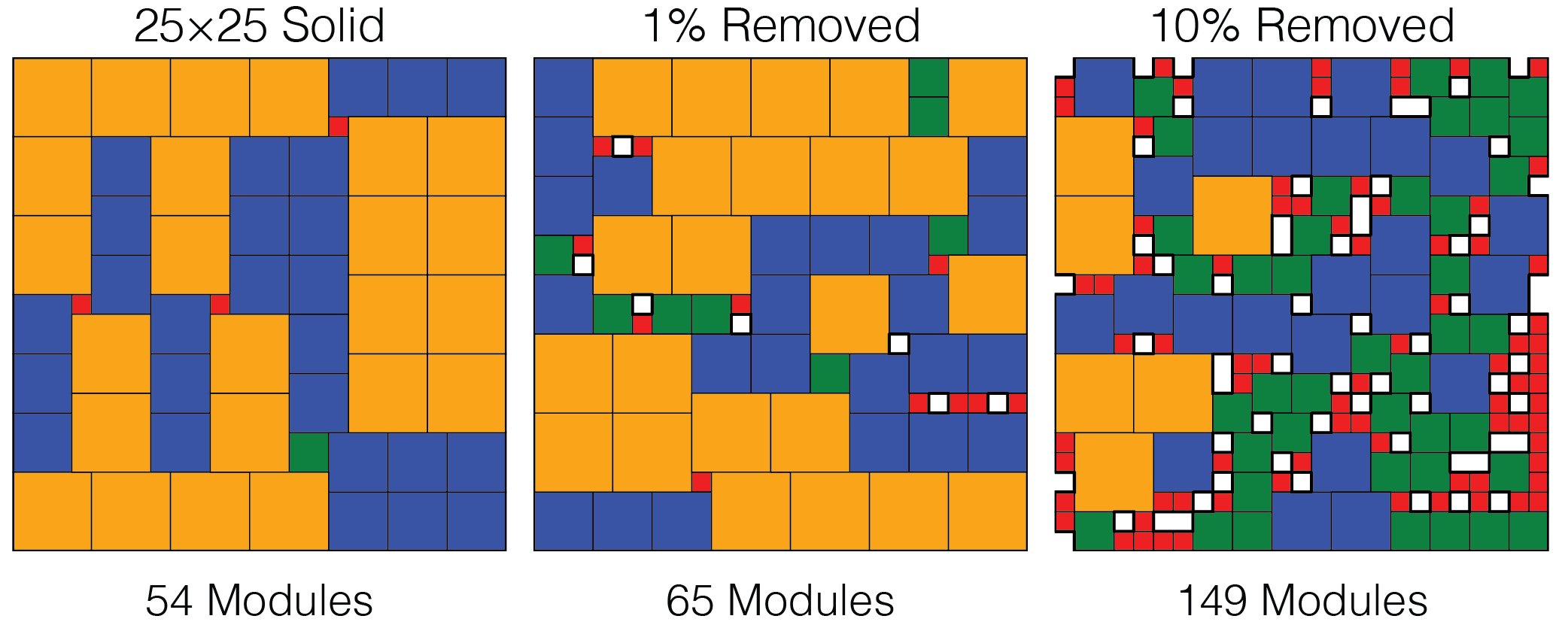}
\caption{\label{fig:decomposition}
Decompositions for panels with increasing numbers of random cutouts, with modules colored by size.
A 25×25 square panel (left) solves in 3.91 seconds. Removing just 1\% of units (center) more than doubles the runtime to 8.68 seconds, while removing 10\% (right) dramatically reduces it to 0.46 seconds.}
\Description{Decomposition runtime across varying panel irregularities. (A) Regular panel: 3.9~s solve. (B) 1\% holes: runtime doubles. (C) 10\% holes: runtime drops to 0.46~s. Demonstrates irregularity impact on the ILP solver.
}
\end{figure}

For patterns with regular panels (e.g., the square panel in Fig.~\ref{fig:decomposition}, left), runtime tends to grow exponentially with size, with occasional spikes in difficulty. Introducing small irregularities—such as randomly removing 1\% of units (Fig.~\ref{fig:decomposition}, center)—can significantly increase runtime. However, more extensive removal (e.g., 10\%) can lead to much faster solves (Fig.~\ref{fig:decomposition}, right), likely due to the resulting smaller contiguous regions reducing the number of feasible module placements and, in turn, the number of ILP variables.

For patterns composed of multiple panels, runtime scales approximately linearly with the number of components, as each panel can be solved independently. These findings suggest that preprocessing large patterns into smaller, disconnected regions can improve predictability and help avoid worst-case exponential runtimes.

\section{Fabrication Details}\label{app:fabrication}

\begin{figure}[h]
\includegraphics[width=\columnwidth]{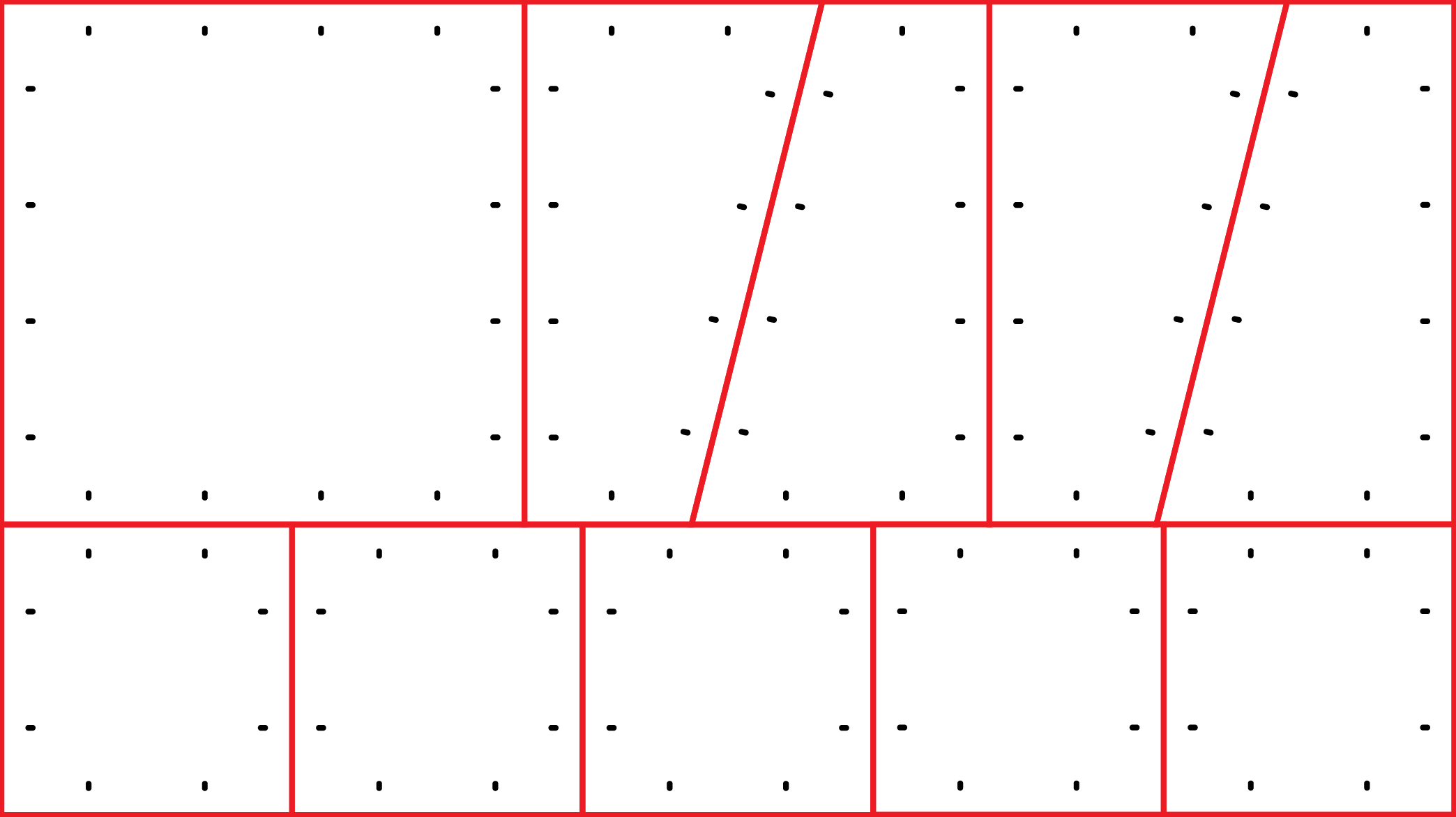}
\caption{\label{fig:fabrication}Dart modules can be packed into rectangles by reflecting one of their trapezoidal halves. Red lines denote module outlines, and black ticks mark connector positions.}
\Description{Fabric layout for module cutting. Red outlines show rectangular and mirrored dart modules; black dots mark connector positions.}
\end{figure}

Before cutting, the fabric was ironed to remove wrinkles. Modules were compactly arranged to minimize material waste. For example, \figref{fig:fabrication} shows an efficient layout of dart modules—packed as rectangles—and foundation modules. We used a Universal 120W laser cutter for fabrication. For light cottons, we used 50\% power and 100\% speed for cutting, and 12\% power at 100\% speed for optional engraving of markers. For denim, we used up to 100\% power and 80\% speed. These settings were tuned to avoid cutting completely through the material—preventing loose pieces during the process—while still allowing for easy removal after cutting.

\end{document}